\documentclass{aa}

\usepackage{graphicx}
\usepackage{txfonts}
\usepackage{natbib}
\bibpunct{(}{)}{;}{a}{}{,} 

\usepackage[colorlinks=true]{hyperref}

\newcommand{\rms}{\mathrm{rms}}

\def\mathLarge#1{\mbox{\large $#1$}}

\begin{document}
  
  \title{Dynamics of dust grains in turbulent molecular clouds} \subtitle{Conditions for decoupling and limits of different numerical implementations.}
   \author{B. Commerçon
           \inst{1}
           ,
         U.~Lebreuilly \inst{2,1},
            D. J.~Price\inst{3},
            F.~Lovascio\inst{1},
         G.~Laibe\inst{1},
            and
            P.~Hennebelle\inst{2}
          }
   \offprints{B. Commerçon}

   \institute{ Univ Lyon, Ens de Lyon, Univ Lyon1, CNRS, Centre de Recherche Astrophysique de Lyon UMR5574, F-69007, Lyon, France\\
             \email{benoit.commercon@ens-lyon.fr}
    \and
Universit\'{e} Paris-Saclay, Universit\'{e} Paris Cité, CEA, CNRS, AIM, 91191, Gif-sur-Yvette, France
              \and
              School of Physics and Astronomy, Monash University, Clayton, Victoria 3800, Australia
          }

   \date{Received October 10th 2022; accepted December 17th 2022}

 \abstract{Dust grain dynamics in molecular clouds is regulated by its interplay with supersonic turbulent gas motions. The  conditions under which  interstellar dust grains decouple from the dynamics of gas in molecular clouds remain poorly constrained.}   
{We first aim to investigate the critical dust grain size for dynamical decoupling, using both analytical predictions and numerical experiments.  Second, we aim to set the range of validity of  two fundamentally different numerical implementations for the evolution of dust and gas mixtures  in turbulent molecular clouds.}
{We carried out a suite of numerical experiments using two different  schemes to integrate the dust grain equation of motion within the same framework. First, we used a monofluid formalism (or often referred to as single fluid) in the terminal velocity approximation. This scheme follows the evolution of the barycentre of mass between the gas and the dust on a Eulerian grid. Second, we used a two-fluid scheme, in which the dust dynamics is handled with Lagrangian super-particles, and the gas dynamics on a Eulerian grid.}
{The monofluid results are in good agreement with the theoretical critical size for decoupling. We report dust dynamics decoupling for Stokes number $\rm{St}>0.1$, that is, dust grains of $s>4~\mu$m in size. We find that the terminal velocity approximation  is well suited for grain sizes of 10 $\mu$m in molecular clouds, in particular in the densest regions. However, the maximum dust enrichment measured in the low-density material --- where $\rm{St}>1 $ --- is questionable. In the Lagrangian dust experiments, we show that the results are affected by the numerics for all dust grain sizes. At  $\mathrm{St}\ll1$, the dust dynamics is largely affected by artificial trapping in the high-density regions, leading to spurious variations of the dust concentration. At $\mathrm{St}>1$, the maximum dust enrichment   is regulated by the grid resolution used for the gas dynamics. }
{Dust enrichment of submicron  dust grains is unlikely to occur in the densest parts of molecular clouds. Two fluid implementations using a mixture of Eulerian and Lagrangian descriptions for the dust and gas mixture dynamics  lead to spurious dust concentration variations in the strongly and weakly coupled regimes. Conversely, the monofluid implementation using  the terminal-velocity approximation does not accurately capture  dust  dynamics  in the low-density regions, that is,  where $\rm{St}>1$. The results of previous similar numerical work should therefore be revisited with respect to the limitations we highlight in this study.}


\titlerunning{Dynamics of dust grains in turbulent molecular clouds}
\authorrunning{B. Commer\c con et al.}
   \maketitle


\section{Introduction}

The properties of dust grains   regulate the opacity, ionisation, initial conditions of planet formation in star  forming regions, but their evolution remains poorly constrained. Contemporary observing facilities  on the ground and in space have shed light on  dust-grain properties at  various scales from the diffuse interstellar medium (ISM) using PLANCK \citep{planck:11} down to protoplanetary discs  using ALMA and SPHERE \citep[e.g.][]{muro:18}. It is now understood that the dust-grain population evolves at all scales via dynamical interaction with the gas (drag force) and the magnetic fields (Lorentz force), as well as via interaction within the dust-grains population that leads to growth and fragmentation processes \citep[e.g.][]{lesur:22}. The study of the  evolution of dust grains during star and planet formation has therefore attracted significant interest in recent years. 
There is growing evidence from observations and theoretical works that dust-grain properties evolve dramatically within dense cores during the very early stages of star and planet formation, suggesting rapid dust growth and dynamical segregation \citep[e.g.][]{steinacker:10,bate:17, sadavoy:18,galametz:19,lebreuilly:20,guillet:20,tsukamoto:21}. It remains unclear as to whether or not dust grains also evolve at larger scales within molecular clouds, prior to protostellar collapse. 

Thanks to major advances achieved in computational astrophysics for modelling dust and gas mixtures  as well as in hardware development, it is now possible to study the dynamical evolution of dust grains at various scales in the ISM with a number of astrophysical codes \citep[see][for a review]{teyssier:19}. In particular, recent works focused on the dynamics of dust grains within molecular clouds and report variation of the dust concentration \citep{hopkins:16,tricco:17,mattsson:19a}. This result is of significant importance because it can have different astrophysical implications. First, the dust-to-gas ratio could vary in molecular clouds beyond the canonical value of 1 percent. Second, dynamical decoupling of grains of different sizes could favour  dust growth and fragmentation processes.  Variations in dust-grain concentration and size distribution will then have potential implications on both heating and cooling, as well as  on mass estimates from observations through opacity variations \citep{ormel:11}.  Furthermore, the non-ideal magnetohydrodynamics resistivities within collapsing dense cores, which regulate the formation of protoplanetary discs, are highly sensitive to dust-grain size distribution \citep{guillet:20,marchand:21}.

However, the critical dust-grain size for measurements of dynamical decoupling varies between these latter studies. \cite{hopkins:16} report significant dust decoupling (variation in dust concentration by more than a factor 1000) for dust grains of $s_\mathrm{grain}\geq 0.01~\mu$m in size. Similarly, \cite{mattsson:19a} report dust-ratio variations for sizes $s_\mathrm{grain}> 1~\mu$m. On the contrary, \cite{tricco:17} report significant decoupling only for dust grains larger than $ 10~\mu$m. In addition,  \cite{mattsson:19b} find that very small grains (typically nanometer size) are also clustering, which increases the local grain density by at least a factor of a few in their models. 

At first glance, the difference in the critical size for dynamical decoupling might seem insignificant, but it can have dramatic effects on the evolution of dust-size distribution in the ISM.  The dust-grain size distribution in the diffuse ISM is relatively well constrained, with maximum sizes of $<1$ $\mu$m \citep{guillet:18}  and a peak of the distribution at around $0.1-0.3~\mu$m  in the Milky Way \citep{weingartner:01,draine:03,guillet:18}. This peak  lies just in the range where  \cite{hopkins:16} and \cite{mattsson:19a} report decoupling while  \cite{tricco:17} do not.

The cause of the discrepancy in the literature 
 for the critical size for decoupling remains unknown. Both the physical and numerical setup are different from one study to another. In the following, we investigate whether these discrepancies may arise from differences in numerical implementation. We briefly present the current state of the art in the numerical methods used, as well as their main limitations, which we will test in this study.

\cite{hopkins:16} and \cite{mattsson:19a} use a two-fluid model, where the dust and the gas are considered as two separate fluids interacting via a drag force. These authors also use two different fluid descriptions. In \cite{mattsson:19a}, the gas is handled on a uniform Cartesian grid (Eulerian approach), while the dust fluid is handled using inertial particles (Lagrangian approach). \cite{hopkins:16} use a mesh-free method  where the gas and the dust fluids are handled using two separate particle distributions \citep{hopkins:15}. In both implementations, the back-reaction of the dust onto the gas is not considered.

In addition, very small grains, which are well coupled to the gas, can be assimilated to gas tracer particles; that is, they mostly move with the gas velocity (with a tiny velocity shift).  \cite{price:10} and \cite{cadiou:19} showed that classical particle-integration schemes based on velocity increments interpolated from the grid (using for instance a Cloud-In-Cell (CIC) algorithm) cannot properly trace the gas dynamics. These authors show that the velocity tracers aggregate in converging flow (high-density region) leading to artificial clustering. The density in the vicinity of converging regions can therefore be overestimated by one order of magnitude, while it is largely underestimated around filaments. We test whether this result affects the dynamics of very small dust grains treated as Lagrangian particles by comparing their distribution to the distribution of dust velocity tracer particles.

On the other hand, \cite{laibe:14a,laibe:14b} propose a  full monofluid approach, which is well adapted for dust grain dynamics as long as the fluid approximation remains valid (Stokes number ($\mathrm{St}$) of the order unity). The  full monofluid system of equations for the gas and dust dynamics and the system  of equations of the two-fluid formalism are mathematically equivalent. The monofluid approach consists of a change of variables, which allows us to follow the evolution of the barycentre of the mixture   and the relative velocity difference and concentration of the dust.  We have further simplified this formalism using the so-called diffusion approximation \citep{price:15}, which is based on the terminal velocity approximation \citep[TVA,][]{youdin:05}. In this approximated monofluid formalism, the drift velocity is set directly by the force budget on the gas and the dust, and only one equation (the mass conservation) needs to be solved  per dust size. In the context of molecular clouds, this diffusion approximation has been used in \cite{tricco:17} and is well suited for well-coupled grains with $\mathrm{St}<1$; it is not valid for larger grains, because the velocity difference between the dust and the gas is underestimated.

In this study, we perform a comprehensive test of well-controlled numerical experiments which allows the use of both the monofluid formalism in the diffusion approximation and a two-fluid method based on inertial Lagrangian particles for the dust. We focus on the effect of the numerical implementations on the dust dynamics in driven dusty turbulence experiments. We will explore the effect of the physical parameters (amplitude and properties of turbulence and the effect of magnetic fields) in a forthcoming study. The paper is organised as follows. In Section 2, we estimate the expected critical dust grain size for decoupling within typical turbulent molecular clouds from simple analytical arguments. The numerical methods and the physical setup are described in Section 3. Section 4 is devoted to the results obtained with the monofluid implementation. We then compare with the two-fluid results in Section 5. In Section 6, we discuss the limitation of this work and the comparison with previous works.

\section{Condition for decoupling in molecular clouds}

\subsection{Turbulence in molecular clouds}
In a turbulent cloud of size $L$ and internal velocity dispersion $v_\rms$, the dynamical time is defined as the time at which a turbulent fluctuation travels through the blob:
\begin{equation}
t_\mathrm{turb} \equiv \frac{L}{v_\rms},
\end{equation}
where $t_\mathrm{turb}$ can been seen as the time needed for the turbulence to completely renew all the hydrodynamic fields of the flow. In molecular clouds, observations show that the amplitude of the turbulence $v_\rms $  and the gas density $n$ scale with the size $L$ as \citep{larson:81,heyer:04,roman:11,hennebelle:12}

\begin{eqnarray}
v_\rms &\simeq&1~\mathrm{km s}^{-1} \left(\frac{L}{1~\mathrm{pc}} \right)^{0.5},\nonumber \\
n  &\simeq& 3000~ \mathrm{cm}^{-3} \left(\frac{L}{1~\mathrm{pc}} \right)^{-0.7}.
\label{eq:Larson}
\end{eqnarray}
We aim to study dust dynamics in typical conditions that satisfy the two aforementioned scaling relations. 

\subsection{Dust grain and gas dynamical (de)coupling\label{sec:scrit}}
Dust grains experience a drag force from the collisions with gas particles. The drag force experienced by a dust grain can be written as 
\begin{equation}
\mathbf{F}_\mathrm{g} = \frac{m_\mathrm{grain}}{t_{\mathrm{s,grain}}}\Delta \mathbf{v},
\end{equation}
where $\Delta \mathbf{v} \equiv \mathbf{v}_\mathrm{g} - \mathbf{v}_\mathrm{d} $ is the differential velocity between the dust grain and the gas (locally described by a Maxwellian velocity distribution), and $t_{\mathrm{s,grain}}$ is the grain stopping time, which characterises the time needed for the dust grain to adjust its velocity to a change of gas velocity.  For typical physical conditions of molecular clouds, the mean free path of the gas is much larger than the size of spherical dust grains; this is the Epstein regime \citep{epstein:24}. In this regime,  the stopping time is defined as 

\begin{equation}
t_{\mathrm{s,grain,0}} = \sqrt{\frac{\pi \gamma}{8}} \frac{\rho_\mathrm{grain}}{\rho_\mathrm{g}}\frac{s_\mathrm{grain}}{c_\mathrm{s}},
\label{eq:ts}
\end{equation}

where $\rho_\mathrm{grain}$ is the grain intrinsic density \citep[typically $1-3$ g cm$^{-3}$,][]{love:94}, $s_\mathrm{grain}$ is the grain size, $\gamma$ the heat specific ratio of the gas, $\rho_\mathrm{g}$ the gas density, and $c_\mathrm{s}$ the gas sound speed.
 In molecular clouds, the flow is supersonic, as described above. As a consequence, the  velocity drift between the dust  and the gas can exceed the sound speed locally. In that case, we  apply a correction for supersonic flow to handle this regime and the stopping time is defined as  \citep{kwok:75,draine:79}
\begin{equation}
        t_{\mathrm{s,grain}} = t_{\mathrm{s,grain,0}} \left( 1+\frac{9\pi}{128} \frac{\Delta \bf v^2}{c_\mathrm{s}^2}\right)^{-1/2}.
        \label{eq:ts_kwok}
\end{equation}
Assuming that all grains have the same intrinsic density, the larger the grain, the longer the stopping time, and consequently, the smaller grains are the most coupled to the gas. The degree of coupling between the dust grain and the gas is characterised by the Stokes number, which is defined as

\begin{equation}
\mathrm{St} = \frac{t_\mathrm{s}}{t_\mathrm{dyn}},
\label{Eq:Stokes}
\end{equation}
where  $t_\mathrm{dyn}$ is the typical dynamical time of the system and $t_\mathrm{s} \equiv \frac{\rho_\mathrm{g}}{\rho}  t_{\mathrm{s,grain}} $ is the fluid stopping time of the dust species (hereafter stopping time), $\rho$ being the total density of the gas and the dust. For $\mathrm{St} \ll 1$, the gas and the dust are dynamically coupled. 
\subsection{Critical grain size for dynamical decoupling}

In turbulent molecular clouds, the dynamical time is the turbulent time $t_\mathrm{turb}$. One can consider that a dust grain decouples from the gas dynamics if $\mathrm{St} >1$, which yields the critical dust grain size (ignoring the correction for supersonic flow in the stopping time expression):
\begin{equation}
s_\mathrm{crit} = \sqrt{\frac{8}{\pi \gamma}}\frac{\rho_\mathrm{g} L}{\rho_\mathrm{grain}\mathcal{M}}.
\end{equation}

According to the Larson relations (\ref{eq:Larson}), the critical dust grain size scales as
\begin{equation}
s_\mathrm{crit}\sim 40~\mu\rm{m} \left( \frac{\rho_\mathrm{g}}{10^{-20}~\rm{g}~\rm{cm}^{-3}}\right)\left(\frac{L}{1~\rm{pc}}\right)\left( \frac{\rho_\mathrm{grain}}{1~\rm{g}~\rm{cm}^{-3}}\right)^{-1} \left(\frac{\mathcal{M}}{10}\right)^{-1} ,
\label{eq:scrit}
\end{equation}
where  $\mathcal{M} = v_\mathrm{rms}/c_\mathrm{s}$ is the turbulent Mach number, and we have taken $\gamma=5/3$. Dust grains of a few 10s microns are therefore expected to decouple significantly  from the gas dynamics within a crossing time $t_\mathrm{turb}$. Numerical experiments show that the decoupling  is already significant for Stokes numbers $\mathrm{St} \geq 0.1$ \citep{dipierro:15,lebreuilly:20}, which indicates that dust grains $>1~\mu$m are expected to decouple  on a dynamical timescale as well.

This value of $s_\mathrm{crit}$ is consistent with the maximum dust-grain sizes expected in molecular clouds \citep[$<1~\mu$m, e.g.][]{kohler:15}. Recent works have shown that grain growth could be at play in the dense and turbulent ISM \citep{ormel:09,guillet:20}. In this study, we therefore consider grains up to sizes of $10~\mu$m.

\section{Numerical methods and initial conditions}
\subsection{The \ttfamily{RAMSES}\rm~code}

We use the adaptive-mesh-refinement (AMR) code \ttfamily{RAMSES }\rm\citep{teyssier:02} which integrates the Euler equations in their conservative form using a second-order finite volume Godunov scheme. In this work, we do not take into account magnetic fields or gravity. In addition, we use a uniform grid, which is well suited to the study of turbulent boxes. 

Our setup follows the classical one designed for investigating compressible turbulence in molecular clouds using turbulent boxes. The boundary conditions are periodic for all hydrodynamical quantities. The gas thermal evolution is isothermal. We use the HLL Riemann solver combined  with a {\it minmod} slope limiter for the hyperbolic part and {\it upwind} for the dust solver \citep[see e.g.][for more details]{toro:99}. In our experience, this combination of solver and slope limiter is the best compromise  between accuracy and robustness.

 We employ the same module to drive the turbulence as that used by \cite{commercon:19}. This latter is based on a source term, a force, in the momentum equation. The turbulent force field is generated in Fourier space  and its mode is given by a stochastic method based on the Ornstein-Uhlenbeck process   \citep{eswaran:88,schmidt:06}. For more details on the turbulence forcing module, we refer readers to the work of \cite{schmidt:09} and \cite{federrath:10} on which our implementation is based. We set the forcing to a mixture of compressive and solenoidal modes, with a compressive force power equal to one-third of the total forcing power. The turbulence is driven on the scale of the computational box, with a peak at half the box length. 

\subsection{Dust and gas dynamics implementations}

\subsubsection{Monofluid, two-fluid, and multi-fluid: terminology explained}
The terms monofluid, two-fluid, Lagrangian dust, and Eulerian dust are often used in this paper, as well as much of the literature in the field. Therefore, it is useful to clarify what is meant by these terms in this context. The two formulations of dust hydrodynamics, `two-fluid' or `multi-fluid' and `monofluid', differ fundamentally in one aspect, namely in how the equations for the dust-gas mixture are written. In a multi-fluid approach, dust species are treated as additional fluids with their own Navier-Stokes equations coupled to the gas by a drag term. The monofluid description is instead designed to describe the mixture as a single fluid, with velocity equal to the barycentre velocity of the mixture, and evolving the drift velocity, which is the vector field that describes the velocity difference between dust and gas. The two descriptions are mathematically equivalent; however, starting from the monofluid description, it is possible to construct approximate descriptions of the mixture by truncating the drift velocity expansion at different orders. The monofluid method in this paper is a TVA of the complete monofluid equations (see section ~\ref{Sec:implemMono}) . A further level of complexity is added to the problem when one considers that, when constructing a multi-fluid solver, it is not required that the same approach be used to solve all the fluid equations (provided the drag term is treated correctly). Hybrid schemes solve the gas dynamics on a Eulerian grid, but solve the dust dynamics using Lagrangian super-particles. This approach may provide some advantages, especially in the limit of $\mathrm{St} \to \inf$ where the dust particle dynamics becomes ballistic mechanics. The two-fluid scheme in this paper is one such `hybrid scheme', which solves the gas dynamics on a grid, but solves the dust dynamics by integrating the motions of Lagrangian dust particles.

\subsubsection{Monofluid implementation}\label{Sec:implemMono}

The implementation in \ttfamily{RAMSES }\rm of dust dynamics in the diffusion approximation ---which is based on the TVA---  and the monofluid formalism was presented in \cite{lebreuilly:19}. We use the multi-species framework, which allows us to follow the coupled dynamical evolution of $\mathcal{N}_\mathrm{d}$ different grain species on a single numerical simulation. Below, we describe the main characteristics of our monofluid implementation. We refer readers to  \cite{youdin:05}, \cite{laibe:14b,laibe:14a}, \cite{price:15}, \cite{hutchison:18}, and \cite{lebreuilly:19} for more details on the derivation of the diffusion approximation in the monofluid formalism.

We first define the monofluid hydrodynamical quantities. The total mixture density $\rho$ is
\begin{equation}
\rho\equiv\rho_\mathrm{g}+\sum_{k=1}^{\mathcal{N}_\mathrm{d}}\rho_{\mathrm{d},k},
\end{equation}
where $\rho_{\mathrm{d},k}$ is the dust fluid density of the $k^{\mathrm{th}} $ size bin. The total dust density is simply $\rho_{\mathrm{d}}=\sum_{k=1}^{\mathcal{N}_\mathrm{d}}\rho_{\mathrm{d},k}$. 
The mixture barycentric velocity $\bf{v}$ is the defined as
\begin{equation}
\bf{v} \equiv \mathLarge{\frac{\rho_\mathrm{g} \bf{v}_\mathrm{g}+\sum_{k=1}^{\mathcal{N}_\mathrm{d}}\rho_{\mathrm{d},k}\bf{v}_{\mathrm{d},k}}{\rho_\mathrm{g}+\rho_\mathrm{d}}},
\end{equation}
where $\bf{v}_{\mathrm{d},k}$ is the velocity of the $k$ bin dust fluid. The dust ratio of dust species $k$ is then $\epsilon_k\equiv\rho_{\mathrm{d},k}/\rho$ and the total dust ratio is $\mathcal{E}={\rho_\mathrm{d}}/\rho$ \footnote{The dust ratio is the  ratio between the local dust density  and the local total density (gas and dust mixture), which must not be confused with the dust-to-gas ratio, which is the mass ratio between the dust and the gas.} . 

The dust and gas mixture dynamical coupling is characterised by the stopping time 
\begin{equation}
t_{\mathrm{s},k}\equiv  \sqrt{\frac{\pi \gamma}{8}} \frac{\rho_{\mathrm{grain},k}}{\rho}\frac{s_{\mathrm{grain},k}}{c_\mathrm{s}}=\frac{\rho_\mathrm{g}}{\rho}t_{\mathrm{s,grain}},
\end{equation}
which leads to the effective stopping time $T_{\mathrm{s},k}$ in the case of multiple dust species. This coupling accounts for the interaction between dust species due to their cumulative back-reaction on the gas:
\begin{equation}
T_{\mathrm{s},k}\equiv \frac{t_{\mathrm{s},k}}{1-\epsilon_k} - \sum_{l=1}^{\mathcal{N}_\mathrm{d}}\frac{\epsilon_l}{1-\epsilon_l}_{\mathrm{s},l}.
\end{equation}
Finally, we introduce the mean stopping time,
\begin{equation}
\mathcal{T}_\mathrm{s}\equiv \mathcal{E}\sum_{k=1}^{\mathcal{N}_\mathrm{d}}\epsilon_kT_{\mathrm{s},k}.
\end{equation}

The set of equations of the evolution of the gas and dust mixture is then
\begin{eqnarray}
\frac{\partial \rho}{\partial t} &+& \nabla\cdot \left[\rho\bf{v} \right] =  0, \nonumber \\
\frac{\partial \rho \bf{v}}{\partial t} & + &\nabla \cdot\left[\rho \mathbf{{v}} \otimes \mathbf{{v}} + P_\mathrm{g} \mathbb{I}  \right] = \rho \bf{f}\rm ,\nonumber\\
\frac{\partial \rho_{\mathrm{d},k}}{\partial t} &+& \nabla\cdot \left[\rho_{\mathrm{d},k}\left(\bf{v}\rm+\frac{T_{\mathrm{s},k}\nabla P_\mathrm{g}}{\rho}\right) \right] =  0,~~\forall k \in [1,\mathcal{N}_\mathrm{d}],
\label{eq:solver}
\end{eqnarray}
where $P_\mathrm{g}$ is the gas pressure, and $\bf f$ is the acceleration applied to account for the turbulence driving. 
The back-reaction of the dust onto the gas is taken into account in this monofluid framework. We note that no energy equation is considered here, because we use the isothermal approximation.

This set of equations is integrated using the classical second-order Godunov scheme of \ttfamily{RAMSES}\rm, which is complemented by the second-order scheme developed by \cite{lebreuilly:19} to integrate the dust differential dynamics.
We refer readers to \cite{lebreuilly:19,lebreuilly:20} for more details on the numerical implementations, as well as tests and applications to prestellar dense core collapse and protostellar disc formation.

\subsubsection{Two-fluid implementation}
We developed an  implementation to account for the dynamics of inertial Lagrangian super-particles in the two-fluid formalism for this study. Our implementation is based on the particle-mesh method used for  tracer particles already implemented in \ttfamily{RAMSES}\rm. The method is well tested, and is used, for example, to run large chemical networks as postprocessing, as it stores the density and temperature of individual parts of the fluid \citep[e.g.][]{coutens:20}.

The dust fluid here is represented by Lagrangian super-particles with constant mass moving at the dust velocity $\mathbf{v}_\mathrm{d}$. Their motion is described by
\begin{equation}
\frac{\mathrm{d}\mathbf{v}_\mathrm{d}}{\mathrm{d}t} = \frac{\mathbf{v}_\mathrm{g}-\mathbf{v}_\mathrm{d}}{t_\mathrm{s}}.
\end{equation}

In the case of strong dynamical coupling, the drag term becomes stiff, which leads to prohibitive time-step constrains. We therefore solve the equation of motion analytically, which leads to the integration scheme
\begin{equation}
\mathbf{v}_\mathrm{d}^{n+1} = \mathbf{v}_\mathrm{g}^{n+1} \left(1-e^{-\Delta t/t_\mathrm{s}^{n}}\right) +  \mathbf{v}_\mathrm{d}^{n} e^{-\Delta t /t_\mathrm{s}^{n}}.
\end{equation}

This analytic formulation naturally yields the limits $\mathrm{St\ll 1}$ and $\mathrm{St\gg 1}$. 
The particle velocity is updated  using a CIC interpolation of the gas density and velocity from the grid. The particles are then moved using the second-order midpoint scheme of \ttfamily{RAMSES}\rm~\citep{teyssier:02}. This scheme is very similar to the one used by \cite{mattsson:19a} using the \ttfamily{PENCIL}\rm~code \citep{pencil:21}. As mentioned earlier, with this implementation, the numerical resolution of the dust depends on the number of particles used, but the accuracy of the drag estimate is limited to the grid resolution used for the gas. Finally, this two-fluid implementation does not take into account the back-reaction of the dust onto the gas.

\subsection{Initial conditions and numerical setup}

The initial physical setup consists of a uniform density box, which represents a portion of a large molecular cloud.  We use a uniform grid with resolution ranging from $128^3$ to $512^3$, always assuming periodic boundary conditions. 
The initial density is $\rho_0=4.4\times10^{-21}$~g cm$^{-3}$ and the box length is set to 4 pc, which satisfies the Larson relations (\ref{eq:Larson}). The gas thermal evolution is forced to remain isothermal with a temperature of $T_0=10$~K. In all models, the grain intrinsic density is $\rho_\mathrm{grain}=1$~g~cm$^{-3}$. This value is rather low if one considers that the dust grains are made of a mixture of graphite (2.2 g~cm$^{-3}$) and crystalline
olivine with composition MgFeSiO$_4$ (3.6 g~cm$^{-3}$). Nevertheless, 
we note that at a given Stokes number, the dust size and intrinsic density are degenerate. If we were to chose an intrinsic density of 1~g~cm$^{-3}$ as used in other studies such as \cite{tricco:17}, we would have to divide the dust size by a factor of 3.

In this study, we explore only one level of turbulence in order to focus on the numerical methods. We set the RMS velocity of the dust and gas mixture in accordance with the Larson relation (\ref{eq:ts}), i.e. $v_\rms\simeq 1.9$~km~s$^{-1}$, which corresponds to a Mach number $\mathcal{M}\simeq 10$ . The turbulent crossing time is then $t_\mathrm{turb}\simeq 2.4$ Myr. For the turbulence driving, we use a unique autocorrelation timescale $\mathcal{T}$  (timescale for a full change of the turbulent field), set to one-quarter of the turbulent crossing time, namely $\mathcal{T}\simeq 0.6$~Myr. 

\begin{table*}[tp!]
        \caption{Summary of the simulations parameters (the star labels our fiducial model)}
        \centering
        \begin{tabular}{c c c c c c c c}
                \hline \hline
                Model                                   & $\mathcal{E}_0$& Grid resolution& $\mathcal{N}_\mathrm{d}$ & $N_\mathrm{p}$ & Dust size &  Monofluid & Two fluid \\
                \hline
                $^*$256MRN                &   0.01               &  $256^3$   & 10  &  0& MRN ; $s_\mathrm{grain}\in [1~\mathrm{nm} , 20~\mu\mathrm{m}]$&  \checkmark &  x \\
                128MRN                    &   0.01      &  $128^3$   & 10 &   0& " &  \checkmark & x\\
                512MRN                            &   0.01      &  $512^3$          & 10 &   0& "&  \checkmark & x \\
                256NBR                             &$10^{-5}$&  $256^3$  & 10  &   0& "&  \checkmark & x \\
                256\_2F                  &     $10^{-6}$&         $256^3$ &          5      &         $10^6$      & 1~nm, 0.01~$\mu$m, 0.1~$\mu$m, 1~$\mu$m, 10~$\mu$m&    \checkmark                  &  \checkmark\\
                128\_2F                  &     $10^{-6}$&         $128^3$ &          5      &         $10^6$      &  "  &    \checkmark                 &  \checkmark\\

                128\_2F\_LR                 &     $10^{-6}$&         $128^3$ &          5      &         $125\times 10^3$      &  "  &    \checkmark                 &  \checkmark\\
                128\_2F\_VLR                 &     $10^{-6}$&         $128^3$ &          5      &         $1\times 10^3$      &  "  &    \checkmark                &  \checkmark\\
\hline
        \end{tabular}
        \label{tab:models}
\end{table*}

  In the following, we present two types of numerical simulations. In the first, we identify the critical dust grain size for decoupling using the monofluid solver presented above. In this setup, we have introduced $\mathcal{N}_\mathrm{d}=10$ bins of dust-grain size, ranging from $1~\mathrm{nm}$ to $12~\mu\mathrm{m}$, which corresponds to Stokes numbers ranging from $10^{-5}$ to $0.18,$ respectively. The fiducial resolution is $256^3$. In addition, in Appendix~\ref{Sec:app_BR}, we investigate  the effect of the back-reaction by varying the initial total dust ratio $\mathcal{E}_0$ from $0.01$ (fiducial value) to $10^{-5}$ (a negligible amount of dust should not affect the gas dynamics).  In Appendix \ref{Sec:app_res}, we also propose a resolution convergence study  where the grid size ranges from $128^3$ to $512^3$.  In the second setup, we compare the monofluid results with the ones obtained using the dust Lagrangian particles (2F runs). To this end, we use $\mathcal{N}_\mathrm{d}=5$ bins of dust-grain size ranging from 1 nm to $10~\mu$m. Thanks to our numerical tool, we can compare the two solvers using a single numerical simulation. As our two-fluid formalism does not allow us to account for the back-reaction, we have set the total dust ratio to the very small value of $10^{-6}$ in the monofluid solver. On top of this, we have added $N_\mathrm{p}$ dust Lagrangian super-particles in the two-fluid solver. These particles  do not impact the hydrodynamical fields. Table~\ref{tab:models} summarises the numerical parameters of the various simulations we run in the context of this study.

\subsubsection{Expected limits to the monofluid and two-fluid methods in this setup \label{Sec:expected_limits}}
 From our analysis in Sect.~\ref{sec:scrit}, the expected critical dust-grain size corresponding to $\mathrm{St} =1$ is $s_\mathrm{crit}\gtrapprox 70~\mu$m. As already mentioned, in our experience, grains with Stokes number $\mathrm{St} \geq 0.1$ already decouple significantly on a dynamical timescale. We therefore expect grains with size $s\gtrapprox 7~\mu$m to decouple in our setup.
 


\subsubsection{Maximum dust velocity}

In the low-gas-density regions, the Stokes number can exceed unity even for the smallest grains. In these conditions, the TVA approximation breaks down, leading to large dust velocity. We therefore limit the dust velocity estimated from our monofluid module  to the mean RMS turbulent velocity, that is, $\mathcal{M}c_\mathrm{s}$. We also apply the correction for supersonic flow when the velocity drift between the dust and the gas exceeds the sound speed \citep{kwok:75,draine:79}. 

\subsection{Postprocessing of the grid versus particle data\label{Sec:postprocessing}}

In this study, we analyse sets of data that encompass both Eulerian and Lagrangian variables. Our analysis is based mostly on volume-weighted  probability density functions (PDFs) and density cuts through the computational domain. We use well-tested suites of postprocessing tools, as follows.

For the grid quantities (monofluid),  we simply bin the grid cells according to the value of the density for the PDF and the slices are straightforward. We use the \href{https://github.com/osyris-project/osyris}{\texttt{OSYRIS}}\footnote{\url{https://github.com/osyris-project/osyris}} visualisation package for {\ttfamily{RAMSES}}. For the particle quantities (Lagrangian dust and tracers), we follow the procedure described in \cite{price:10} and \cite{price:11}. First, all particles have the same mass, $m_\mathrm{p}$, which is equal to the total mass of the dust species divided by the number of particles). To produce density maps, we project the particle distribution onto a regular grid, which has the same dimensions as the grid used for the monofluid. We then use the Smoothed-particle hydrodynamics 
(SPH) density calculation routine from the \ttfamily{PHANTOM}\rm~code \citep{price:18}, where the density and smoothing length are iterated self-consistently using classical SPH procedures. We produce cross-section slices of the density field using the \ttfamily{SPLASH}\rm~visualisation software \citep{price:07}. To obtain the volume-weighted PDF from the particles, we follow \cite{price:11} and again interpolate the density on an adaptive grid.

\section{Monofluid results \label{Sec:monofluid}}
\subsection{Time evolution}

\begin{figure}[t!]
        \centering
        \includegraphics[width=0.5\textwidth]{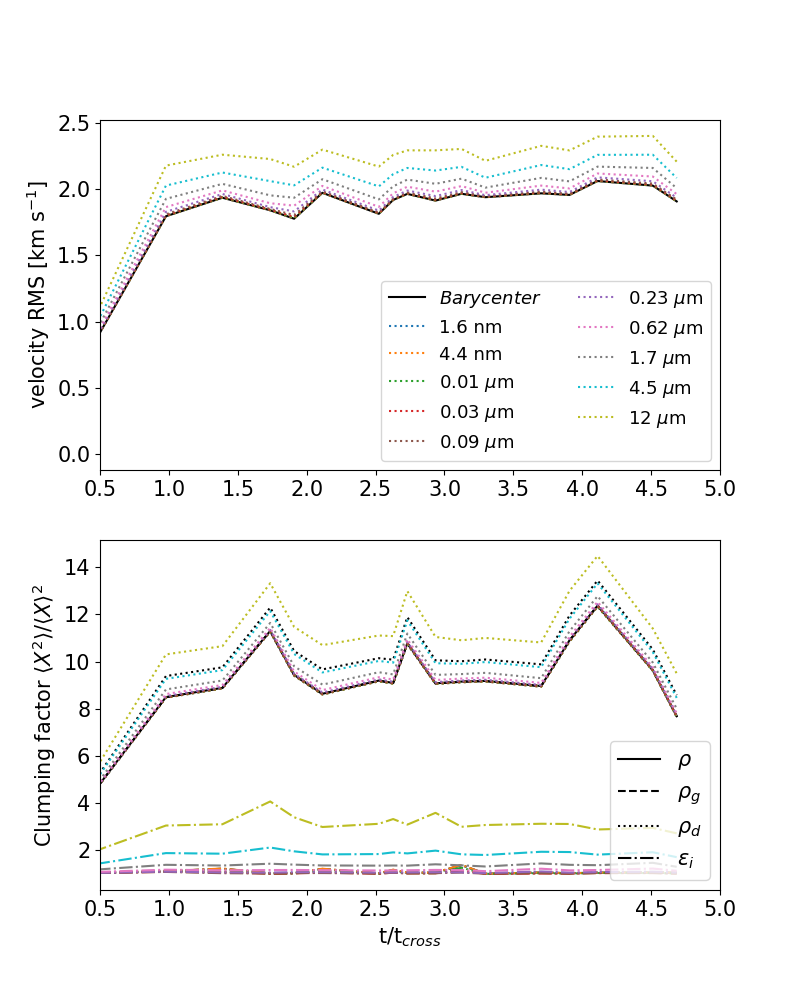}
        \caption{Time evolution of the barycentre and dust-species RMS velocity (top) and of the clumping factor of the density of the different species (bottom) in the fiducial model 256MRN. The time evolution has been normalised according to the turbulent crossing time $t_\mathrm{cross}$. In the bottom panel, the dotted line shows the clumping factor evolution of the total dust density $\rho_\mathrm{d}$ and the dashed-dotted line shows the clumping factor evolution of each dust species concentration $\epsilon_i$. The evolution of the clumping factor  of the gas density (dashed) is indistinguishable from that of the barycentre (solid) density.
        }
        \label{Fig:time_evol}
\end{figure}
\begin{figure*}[t!]
        \includegraphics[width=0.9\textwidth]{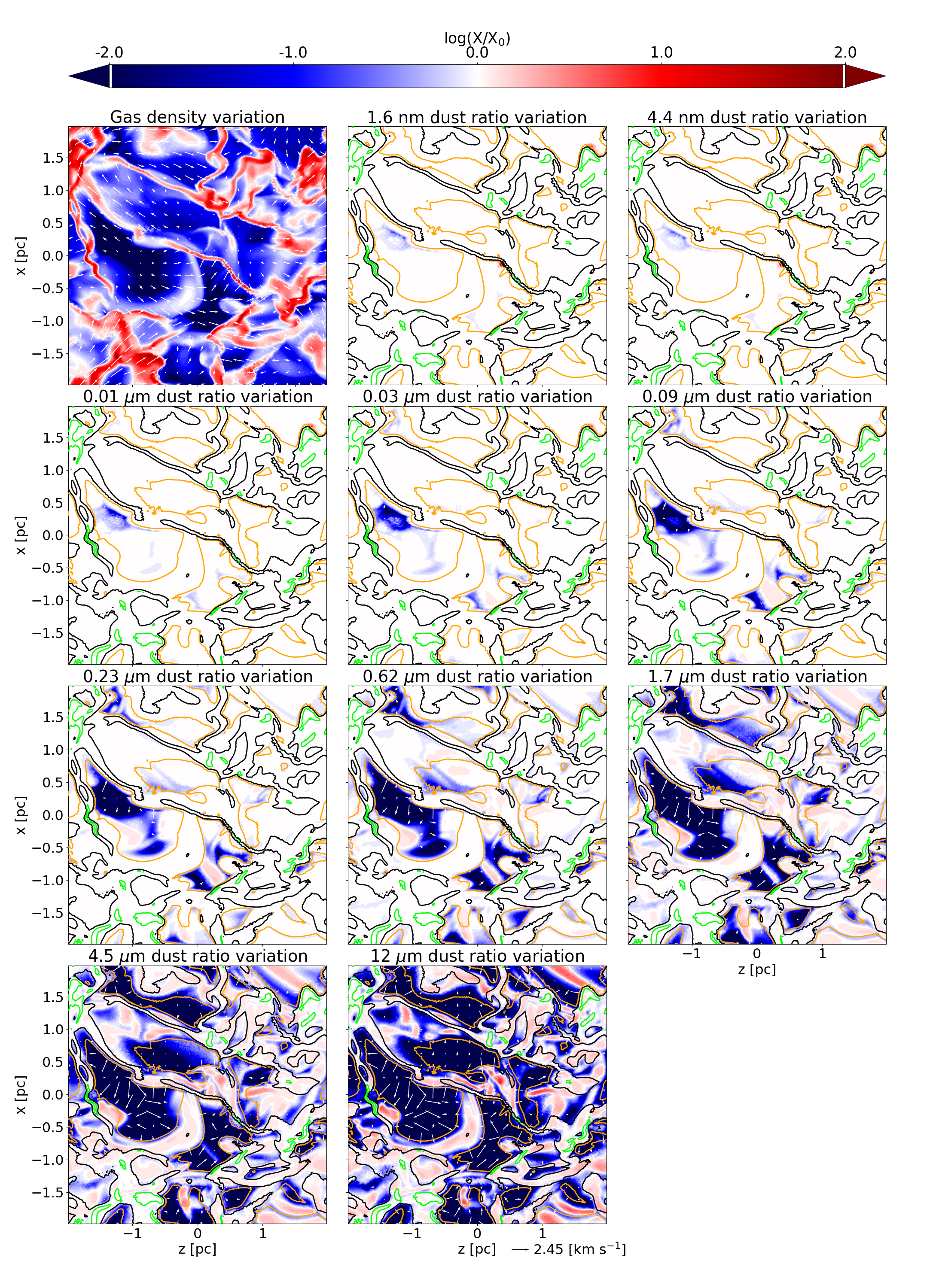}
        \caption{Gas density (top-left) and dust ratio variation maps in the $xz$-plane for the ten dust bins in the 256MRN run. The  variations are given relative to the initial value of the quantity and are shown in logarithmic scale. In the dust-ratio panels, the red colour shows dust-ratio enhancement, while blue indicates a dust-ratio decrease. The isocontours correspond to gas-density variations of $-1$ (orange), 0 (black), and $+1$ (green) in logarithmic scale.  The arrows represent the barycentric velocity (top-left), and the dust-grain drift velocity vectors in the plane for all the other plots. All plots are made at a time corresponding to $3t_\mathrm{cross}$.
        }
        \label{Fig:eps_maps}
\end{figure*}

\begin{figure*}[t!]
        \includegraphics[width=\textwidth]{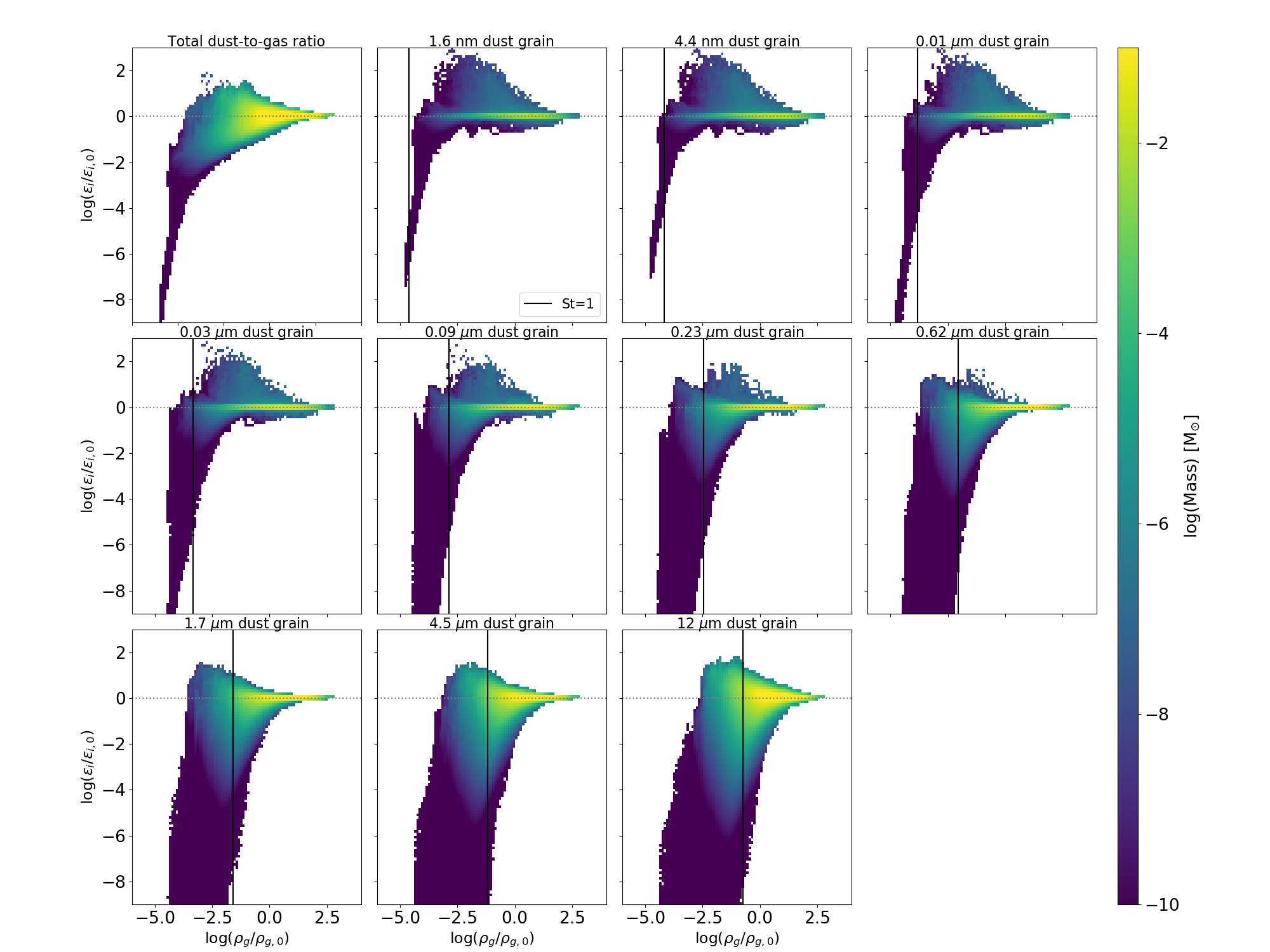}
        \caption{2D histograms of the dust-ratio variations as a function of gas density  for the fiducial model 256MRN. The top-left panel shows the variation in total dust-to-gas ratio $\mathcal{E}$ , while the other panels show the dust-ratio variations of each dust species.  The colour coding indicates the mass of the gas in log scale (in solar units). The horizontal grey dotted line separates the dust-enriched and dust-depleted regions.  The vertical line indicates $\mathrm{St}=1$ (solid)  for each dust species. All quantities are averaged over more than one dynamical time $t_\mathrm{cross}$ and the variations are normalised to the initial value of each dust ratio.
        }
        \label{Fig:enrichement_histo}
\end{figure*}

Figure \ref{Fig:time_evol} shows the time evolution of the RMS velocity $\langle \sqrt{v^2}\rangle$  and of the density clumping factor of the various fluids  (dust, gas, barycentre). The clumping factor $C_X$ of a quantity $X$ is defined as $C_X=\langle{X^2}\rangle/\langle X\rangle^2$. A clumping factor $C_X > 1$ indicates that the field $X$ shows significant variation. The RMS velocity and the density clumping factor of the barycentre and of the ten different dust species are shown. In addition, we show the evolution of the total dust density and gas density clumping in the clumping factor plot. After roughly one turbulent crossing time, the various quantities oscillate around a mean value and do not show any trend as a function of time, which indicates that we have reached a steady state. While dust species with size $s<1~\mu$m do not show any significantly different evolution from the barycentre, the largest dust grains tend to show a decoupling with amplitude in the measured quantities, which remains constant with time. In addition, the fluids consisting of dust grains with sizes $s\geq4~\mu$m have RMS velocities and clumping factors that increase with size. The larger the grain, the larger the turbulence amplitude and the stronger the clumping. Interestingly, the total dust density also shows a stronger clumping than the barycentre and the gas. This is mostly due to the fact that under the MRN size distribution, most of the dust mass is contained in the largest grain species. In summary, this first qualitative analysis shows that grains with $\mathrm{St}>0.1$ decouple from the gas dynamics, in good agreement with our theoretical estimate. In addition, the gas and barycentre quantities exhibit almost identical time evolution, meaning that the gas-phase evolution corresponds to the evolution of the barycentre.

In addition, we verified that  the evolution of the total dust mass contained above given the total density threshold remains roughly constant with time. This indicates that there is no cumulative trapping of the dust grains at any density as a function of time, and that the degree of decoupling also remains constant with time. The quantities we derive in the following are then independent of time, and we can safely analyse our models at a given time snapshot, even though the absolute amplitude of the different quantities might fluctuate with time.

\begin{figure}[thb!]
        \includegraphics[width=0.5\textwidth]{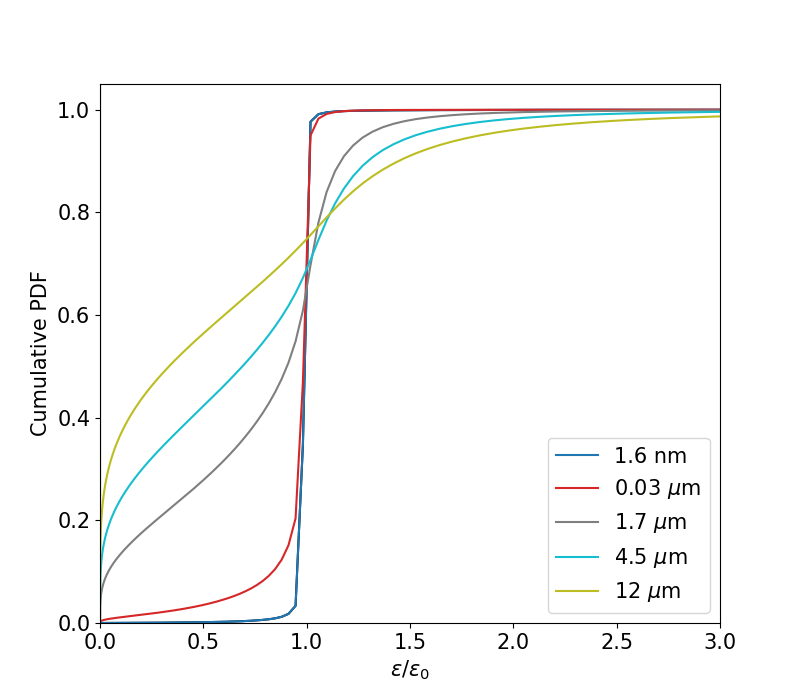}
        \caption{Volume-weighted cumulative PDF of the dust ratio variation for the $1.6$~nm, $0.03$, $1.7$, $4.5,$ and $12~\mu$m dust grains in the fiducial model 256MRN. All quantities are averaged over more than one dynamical time $t_\mathrm{cross}$. 
        }
        \label{Fig:epsilon_pdf}
\end{figure}
\begin{figure}[t!]
        \includegraphics[width=0.5\textwidth]{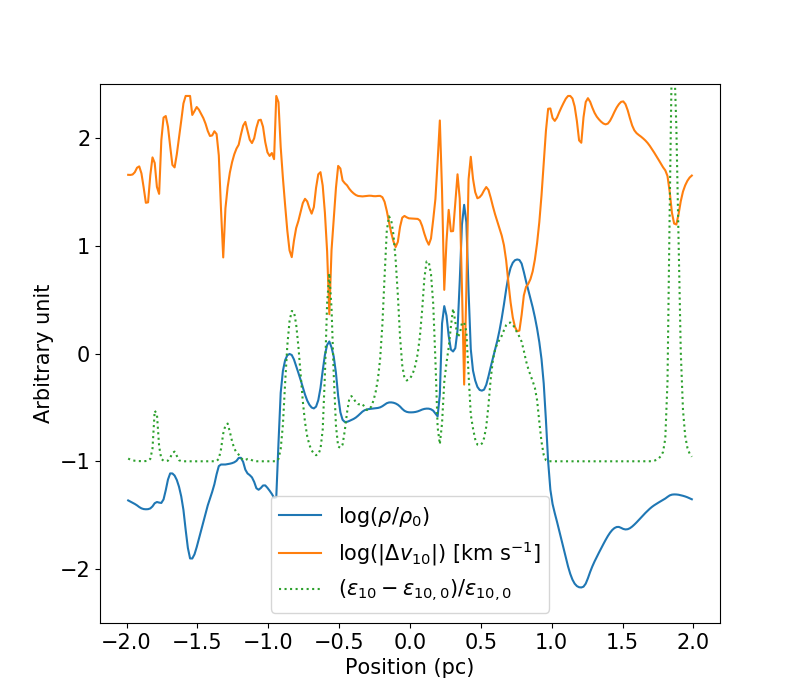}
        \caption{Profile of the total density $\rho$ (blue) and  the drift velocity (orange) and dust ratio (dotted green) for the $12~\mu$m dust grains  in the fiducial model 256MRN. The density is normalised by its initial value. The norm of the drift velocity with respect to the barycentre is normalised to 1 km s$^{-1}$. The variations in dust ratio  are computed with respect to its initial value.  The quantities are plotted against the $x-$axis at a random time ($>2t_\mathrm{cross}$) and  at a random location in the $yz$-plane. Regions with positive relative dust ratio correspond to dust-enriched material.        }
        \label{Fig:profile}
\end{figure}

\begin{figure}[thb!]
        \includegraphics[width=0.5\textwidth]{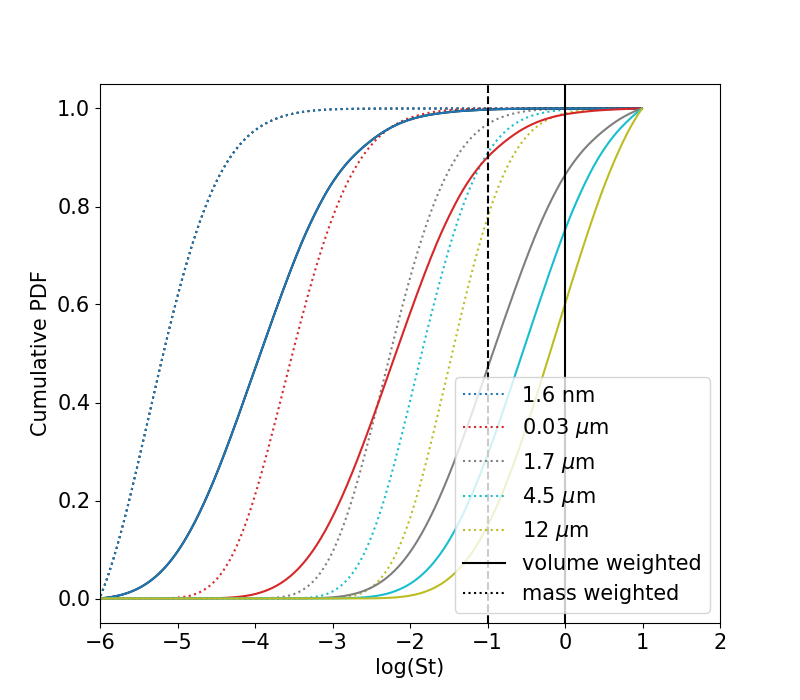}
        \caption{Cumulative PDF of the Stokes number for the $1.6$~nm dust grains and the $0.03$, $1.7$, $4.5,$ and $12~\mu$m dust grains  in the fiducial model 256MRN. The coloured dotted lines show the mass-weighted PDF and the coloured solid lines show the volume-weighted PDFs. The Stokes number is computed using the local gas density and the global turbulent crossing time $t_\mathrm{turb}$.  All quantities are averaged over more than one dynamical time $t_\mathrm{cross}$. The vertical lines indicate the $\mathrm{St}=1$ (solid) and $\mathrm{St}=0.1$ (dashed).
        }
        \label{Fig:stokes_pdf}
\end{figure}

\subsection{Characteristic features of the decoupled regions}

Figure \ref{Fig:eps_maps} shows density maps at time $3t_\mathrm{cross}$ of the variations of the gas density and dust ratio $\epsilon_i$ of each dust species relative to their initial value.  The gas density varies over more than four orders of magnitude with large velocities in the low-density regions. The amplitude of the density variation is typical of isothermal compressible turbulence, with density jumps at shocks varying as $\propto \mathcal{M}^2$.  The amplitude of the dust ratio variations increases with dust-grain size, up to very large values (more than eight orders of magnitude) for the largest grains.  The region that show a large variation in dust ratio, for both depletion and enrichment\footnote{We define the dust enrichment (resp. depletion) as the material where the dust ratio variation is $\epsilon_i/\epsilon_{i,0}>1$ (resp. $\epsilon_i/\epsilon_{i,0}<1$).}, are associated to densities $\rho_\mathrm{g}<\rho_\mathrm{g,0}$ where the largest density gradients develop. As we use an isothermal equation of state, the pressure gradients are proportional to the density gradients. The dust velocity shift is also inversely proportional to $\rho^2$, and so the smaller the density, the larger the velocity drift, which is consistent with our findings.  We also observe strong depletion of dust density for dust grains $>0.1~\mu$m. These depleted regions correspond to low gas densities and therefore to low dust densities. As a consequence, the dust enrichment is mostly at higher densities, because the total dust mass is conserved. 
Interestingly, we observe that the strongly depleted regions (shown in dark blue in Fig. 2,  corresponding to $\log(\epsilon_i/\epsilon_{i,0})<-2$) always sit in low gas densities  where $\log(\rho_\mathrm{g}/\rho_\mathrm{g,0})<-1$. These regions also exhibit the largest dust velocity. On the other hand, the dust-enriched regions do not correspond to higher densities, but rather are found where $-1<\log(\rho_\mathrm{g}/\rho_\mathrm{g,0})<1$. The high-gas-density material shows a nearly uniform dust ratio, which corresponds to the initial value for all dust species, except the largest dust species where variations of a factor of a few can be observed.

In order to better characterise the regions of decoupling, Fig.~\ref{Fig:enrichement_histo} shows the dust-to-gas ratio variations as a function of the gas density for the total amount of  dust, as well as the dust ratio for the various dust-grain species we modelled. All dust species show the largest variations in the dust ratio for gas  density $\rho_\mathrm{g}<\rho_\mathrm{g,0}$. The amplitude of the variations, in particular depletion, increases with dust size. Interestingly, we see that the smallest dust grains ($s<0.1~\mu$m) are  only strongly depleted at the smallest gas density, that is,
$\epsilon_{i}/\epsilon_{i,0}<10^{-2}$, while the enrichment is effective for gas densities  $-4<\log(\rho_\mathrm{g}/\rho_\mathrm{g,0})<1$. The bulk of the mass remains at the initial dust ratio for the smallest grains. These grain species also exhibit Stokes numbers $\mathrm{St}<1$ for the bulk of the mass, which indicates that our monofluid approximation is well suited for these species. 
The  grains  with sizes $>0.1~\mu$m show stronger depletion even at gas densities corresponding to $\rho_\mathrm{g,0}$.  The amplitude of the dust enrichment is lower compared to the smaller grains, typically $\epsilon_{i}/\epsilon_{i,0}<100$, but corresponds to a larger bulk mass. 
In terms of total mass budget, as the largest grains carry the mass, the total dust-to-gas ratio exhibits some variation, but this is only moderate (less than one order of magnitude for the bulk). For grains with sizes $s>1~\mu$m, the maximum dust variation occurs at $\mathrm{St}>1$, which is outside the range of validity of the diffusion approximation. In Sect.~\ref{sec:val_mono}, we show that this is not a problem for the total mass budget because the bulk of the mass is found in the $\mathrm{St}<1$ regions.  However, the amplitude of the dust variations might be inaccurately estimated.

Figure~\ref     {Fig:epsilon_pdf} shows the cumulative volume-weighted PDFs of the dust-ratio variations for various dust species, including the three largest ones. For grains with sizes $s<0.1~\mu$m, the bulk of the volume does not show variation in dust ratio. The dust-enriched region corresponds to less that 5\% of the total volume for the $0.03~\mu$m dust grains. As explained in the previous paragraph, this material corresponds to low gas density. For larger grains, $s>1~\mu$m, more than half of the total volume shows dust-enriched regions. About 4\% of the $12~\mu$m dust-grain material shows dust-ratio variations of greater than a factor of two, and about 1.3\% shows variations of greater than a factor three.

Figure~\ref{Fig:profile} shows the variation of the total density along a line of sight, as well as the drift velocity and dust ratio for the $12~\mu$m dust grains in the fiducial 256MRN model. As our model is isothermal, the density directly traces the pressure, meaning that the density variations correspond to the pressure ones. In the TVA approximation, the drift velocity is proportional to the pressure gradient. The pressure and drift velocity evolutions are anti-correlated. The drift-velocity minima correspond to pressure maxima. The dust-enriched regions are found in the pressure maxima. As already reported, the highest concentrations are measured in the regions of intermediate density ($\rho<\rho_0$). This behaviour confirms that our implementation of the TVA correctly captures the predicted physics.

\subsection{Validity of the diffusion approximation\label{sec:val_mono}}
Figure~\ref     {Fig:stokes_pdf} shows the volume- and density-weighted PDFs of the Stokes number for various dust species, including the three largest ones. The bulk of the mass of all dust species sits in $\mathrm{St}<0.1$ regions. However, the volume-weighted PDFs of the largest grains show large volume fractions with $\mathrm{St}>1$. This corresponds to  $\simeq 60\%$ of the total computational volume for the $12~\mu$m grains, $\simeq 25\%$  for the $4.5~\mu$m grains, and $\simeq 15\%$ for the $1.7~\mu$m grains.   In terms of mass, the fraction of the total dust mass of the largest bin ($12~\mu$m) where the Stokes number is  $\mathrm{St}>1$ is $\simeq 1\%$ ($\simeq 22\%$ for $\mathrm{St}>0.1$).  

We also checked the fraction of the total volume and mass of each dust species where the limitation of the dust velocity is negligible. For the most decoupled dust grains ($12~\mu$m), the fraction of the mass where the velocity floor is active is always $\lessapprox 10^{-6}$, which corresponds to a fraction on the order of a few times $10^{-5}$ of the total volume. This  confirms that the dust diffusion approximation and our numerical implementation are well suited for our purposes. 

\subsection{Summary of the monofluid results}
The main results of our analysis of the monofluid results are as follows:
\begin{itemize}
        \item Dust grains of size $<0.1~\mu$m remain well coupled to the gas.
        \item  Dust grains of size $>1~\mu$m show strong variation in concentration. The decoupling regions correspond mainly to low-density material.
        \item Dust concentrates in the pressure maxima. 
        \item The total dust fraction does not show variation in the high-density regions.
        \item Our monofluid and TVA implementation  is well-suited for the the bulk of the mass (99\% for the largest grains). However, the maximum dust variation for large grain sizes occurs for physical conditions that are  outside the range of validity of the diffusion approximation.
\end{itemize}

\begin{figure}[t!]
        \includegraphics[width=0.5\textwidth]{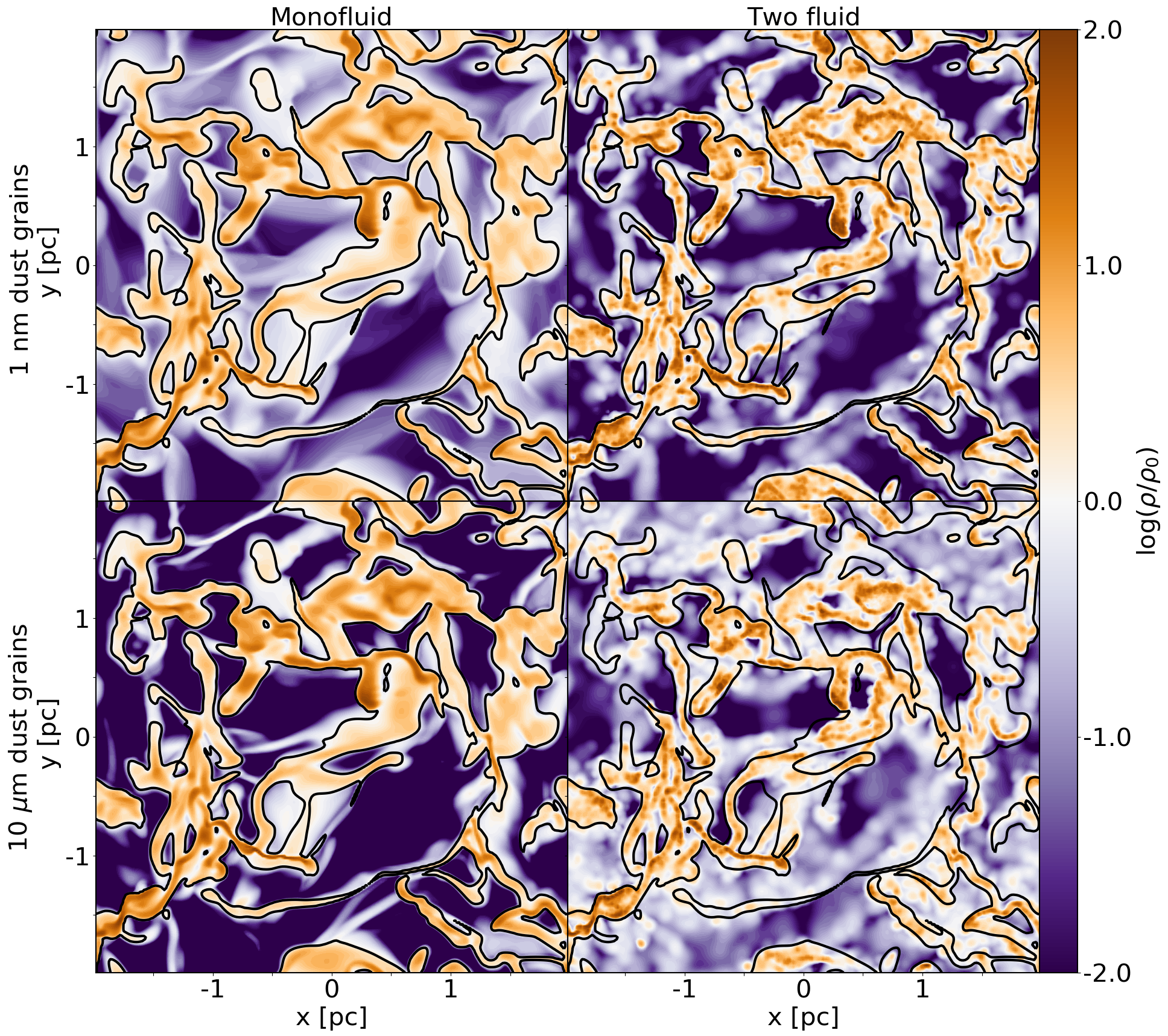}
        \caption{ Density slice in the $z=0$ plane obtained from the 256\_2F run for the 1~nm (top) and 10~$\mu$m (bottom) dust grains at a time corresponding to $\simeq 4.5t_\mathrm{cross}$. The left column represents the density field of the monofluid dust, while the right column represents the density field interpolated from the Lagrangian dust particles. The density is normalised by the initial density $\rho_0$. The black lines represent the contour for which the gas density is equal to its initial value $\rho_0$. }
        \label{Fig:compare_rho}
\end{figure}

\section{Comparison between monofluid and two-fluid formalisms\label{Sec:bifluid}}
In this section, we focus on the comparison of the above results based on the monofluid formalism with those produced using our two-fluid implementation based on Lagrangian superparticles. We also present a  convergence study where we vary both the grid resolution $N$ and the number of particles $N_\mathrm{p}$. 

Our comparison focuses on the evolution of five dust-grain size populations, which are treated  using both the monofluid and the two-fluid solvers in single simulation runs. The dust-grain size varies from 1~nm (well-coupled grains) to $10~\mu$m (least coupled), and the dust-to-gas-mass ratio of each dust species is set to $10^{-6}$. Indeed, as already mentioned, as our two-fluid implementation cannot account for the back-reaction of the dust-grain dynamics on the gas dynamics whereas the monofluid implementation can, choosing a low value of the dust-to-gas ratio means we are mostly neglecting the back-reaction in the monofluid solver.  All other aspects of the monofluid setup are identical to the ones presented in the previous section (except for grain size and total dust mass). In addition to the two solvers, we have  included classical tracer particles designed to trace the dust fluids handled in the monofluid runs, that is, with the velocity computed in the terminal-velocity approximation. The tracer-particle populations are used for comparison with both the monofluid and the two fluid results. 

\subsection{Density fields \label{Sec:bifluid_density}}
\begin{figure}[t!]
        \includegraphics[width=0.5\textwidth]{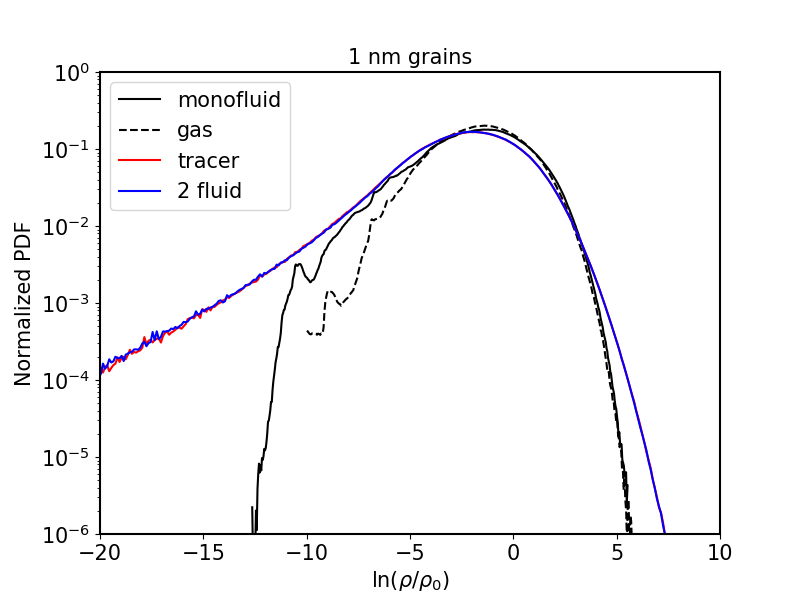}
        \includegraphics[width=0.5\textwidth]{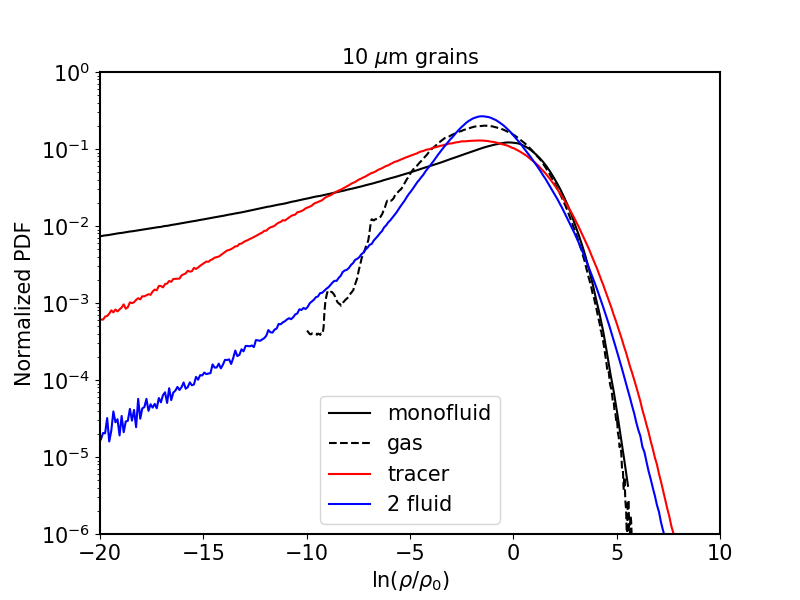}
        \caption{PDFs of the logarithm of density $\ln \rho$  in the 256\_2F run for the 1~nm (top) and 10~$\mu$m (bottom) dust grains. The density variations are normalised to the initial density values for each field. }
        \label{Fig:mean_rho_pdf}
\end{figure}

 Here, we investigate the results of the 256\_2F run. In this run, each dust-grain size is represented by $10^6$ `two fluid' Lagrangian particles and another $10^6$ monofluid `tracer'  particles. Figure \ref{Fig:compare_rho} shows slices of dust-density variations for the 1 nm and 10~$\mu$m dust-grain fluids for both the monofluid and the two-fluid approximations. All quantities are normalised to the initial dust density of each species. We also show the isocontour, which indicates the threshold where the gas density is equal to its initial value. We can therefore compare the dust variations to those of the gas. For the 1~nm dust grains, the dust-density variations from the monofluid closely match the gas-density variations, which is in very good agreement with what we observe in the previous section for the most coupled dust species. The 1~nm dust-grain density obtained with the dust as Lagrangian particles globally matches the density of the gas. The gas-depleted regions correspond to dust-depleted regions. However, one can distinguish small islands of dust-depleted regions within gas-enriched regions and dust-enriched filaments in some gas-depleted regions, which is indicative of decoupling.  For the 10~$\mu$m dust grains, the dust-density variations roughly match the gas variations for the monofluid, with large dust-depleted regions of more than a factor of two. Conversely, the  variations in the dust-density field from the  Lagrangian particles are less important than those found in the  monofluid, with smaller enriched and depleted regions. Large  dust-depleted islands are also observed within the gas-enriched regions. 

\subsubsection{Tightly coupled dust grains (1 nm)}
Figure~\ref{Fig:mean_rho_pdf} shows the averaged PDFs of the density obtained from the gas and the monofluid, the Lagrangian dust particle, and the monofluid tracer-particle populations for the  1 nm and 10~$\mu$m dust grains. We first focus on the PDFs of the 1~nm dust grains. We observed two  clearly distinct behaviours. On the one hand, the gas and monofluid PDFs match relatively well, in particular for the largest densities, while on the other, the tracer and  Lagrangian-dust-particle  PDFs are identical. The dust-tracer-particle PDF does not match the monofluid PDF  at all, while they should match. 

First, comparing the monofluid and the  gas PDFs,  there are more dust-depleted regions as well as gas-depleted ones. As a consequence, there is a slight shift of the peak to high density for the dust population. This is seen in the previous section for tightly coupled dust grains. For the 1~nm dust grains, the expected behaviour would be for the dust density PDFs to be similar to the gas ones, as the gas drag is strongest. However, at low gas density, the Stokes number becomes large enough for the dust to slightly decouple. 

Second, the monofluid tracer particles should trace the monofluid dust species, but as the two PDFs are very different, they do not. We observe an excess  in both  the low- and high-density tails of the tracer particles distribution, meaning that some particles are  trapped in the high-gas-density regions, resulting in a larger depletion in the  regions of low gas density. This result suggests that particles may be subject to artificial clustering, which we investigate in the following subsection. From the identical  tracer- and  Lagrangian-dust-particle population PDFs, we see that   both particle populations experience exactly the same velocity field, computed using either the terminal velocity approximation (monofluid) or our analytical integrator scheme (two-fluid Lagrangian particles). Given the small Stokes numbers of the 1~nm dust grains, the computed velocity is equal to the gas velocity, which further verifies that the dust velocity computed by the monofluid approximation is correct in this regime. These two PDFs depart significantly from the gas-density PDF as well.

\begin{figure}[t!]
        \includegraphics[width=0.5\textwidth]{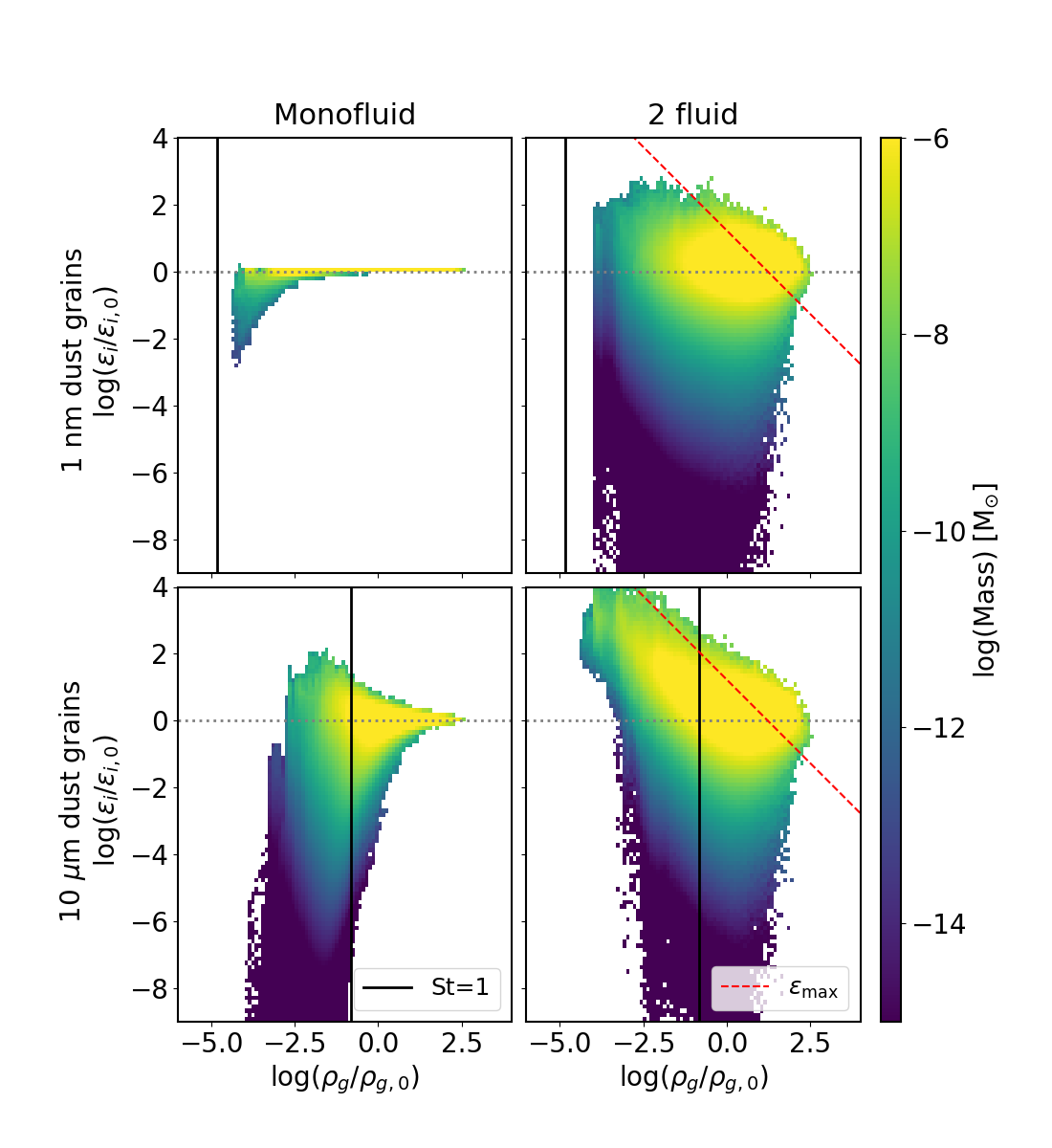}
        
        \caption{Same as Fig. \ref{Fig:enrichement_histo} but for the 256\_2F run. The left column shows the monofluid results and the right column shows the  results when  dust is modelled as Lagrangian particles.  The top row shows the $1$~nm dust grains and the bottom row the  $10~\mu$m dust grains. The red dashed line indicates the maximum dust enrichment as defined in Eq. \ref{Eq:eps_max}. All the material that is above this line has an excessive dust concentration. }
        \label{Fig:dust_enrichment_bifluid}
\end{figure}

\subsubsection{Least-coupled dust grains}
For the 10~$\mu$m dust grains, we observe four different PDFs  in Fig.~\ref{Fig:mean_rho_pdf}. First, the  tracer particles and monofluid PDFs are different, indicating that the tracer particles do not trace the dust fluid as they should for all Stokes regimes. This brings into question the reliability of the results obtained with the classical Lagrangian tracer particles. Further, the monofluid PDF departs from the gas PDF for low density. At high density, where the Sokes number is $\mathrm{St}\ll 0.1$, the two PDFs (gas and monofluid) are a close match, as indicated by the small dust-ratio variations observed previously. The Lagrangian dust-particle PDF falls in the middle of the gas and monofluid PDFs, close to the PDF\ of the tracer particles at high density and to the gas PDF at low density. This last observation seems to indicate that the two-fluid solver based on Lagrangian particles also has problems handling the large gas density corresponding to small Stokes numbers (clustering). The relatively good match with the gas PDF at low density is also questionable, because the dust-grain dynamics should be decoupled from the gas dynamics given the high Stokes numbers. We see in the following sections that this match is not robust, and that the low-density slope of the PDF depends on the number of  particles $N_\mathrm{p}$. 

\subsubsection{Intermediate summary}
The main results of the analysis of the density PDFs are the following:
\begin{itemize}
        \item Two-fluid results exhibit stronger decoupling than the monofluid ones, for all dust sizes.
        \item For small dust-grain sizes, the tracer particles and very small dust Lagrangian super-particles have identical density PDFs.
        \item For large dust-grain sizes, the density PDFs obtained with the monofluid and the two-fluid formalisms do not agree.
        \end{itemize}

\subsection{Dust ratio}

In this section, we  focus on the dust variations  obtained in the 256\_2F run.  We computed the dust enrichment in the two-fluid case with dust as Lagrangian particles by projecting the dust-particle distribution onto a uniform grid of the same resolution as that used for the gas,  and divided by the gas density. For the  dust species handled with the two-fluid solver, we can determine a maximum dust enrichment by assuming that the minimum length scale of the particle distribution cannot go beyond the gas resolution set by the grid for the gas. We define the maximum dust enrichment $\epsilon_\mathrm{max}$ by setting the minimum adaptive resolution length $h_\mathrm{min}$ (see definition in the following section) equal to the grid size $\Delta x$:
\begin{equation}
        \epsilon_\mathrm{max}\equiv\frac{\rho_{\mathrm{d,max}}}{\rho}=\frac{m_\mathrm{p}}{\Delta x^3 \rho}, 
        \label{Eq:eps_max}
\end{equation}
where $m_\mathrm{p}$ is the mass of a particle.

Figure  \ref{Fig:dust_enrichment_bifluid} shows 2D histograms of the dust enrichment obtained in  the 256\_2F run for the monofluid and Lagrangian-particle dust species of 1~nm and $10~\mu$m in size. First, the monofluid results are different from those observed in Fig.~\ref{Fig:enrichement_histo} for the 256MRN run. In particular, there is almost no  dust enrichment for the 1~nm dust grains. In  appendix \ref{Sec:app_res}, we show that this is due to the neglect of the dust back-reaction onto the gas, which mainly affects the dust-enriched tail of the PDFs. The monofluid results for the  $10~\mu$m dust grains are similar to the ones observed in the 256MRN run.

We now compare with the results obtained in the two-fluid case with dust as  Lagrangian particles. For both grain sizes, monofluid and two-fluid solvers give very different results. For the well-coupled dust grains, there is a huge spread in dust enrichment (up to at least two orders of magnitude) in the full density range. As a consequence, the depletion is also large  at all densities. For the $10~\mu$m dust grains, the enrichment is higher at low gas density compared to the monofluid results. In particular, the  regions of lowest gas density are only dust enriched. This corresponds to regions where $\rm St>1$. At high gas density, the enrichment is  similar to that observed for the smallest grains, which indicates that they undergo the same  dynamical (de-)coupling with the high-density gas at low Stokes number. 
The red line in Fig.~\ref{Fig:dust_enrichment_bifluid} indicates the maximum dust enrichment,  which corresponds to the maximum dust density allowed by the gas grid resolution.  The entire region above $\epsilon_\mathrm{max}$ corresponds to dust enrichment with resolution length smaller than the gas resolution, that is, unresolved dynamical coupling. For 1~nm dust grains, only the  regions of high gas density are affected, while for the $10~\mu$m dust grains, the upper part of the distribution follows a trend inversely proportional to the gas density. This suggests that the maximum dust enrichment is regulated by the grid resolution used for the gas dynamics. We observe a maximum enrichment of about four orders of magnitude in the low-gas-density material, which corresponds to the maximum enrichment permitted by the grid resolution used for the gas dynamics. Even if the two-fluid implementation with dust as  Lagrangian particles is better suited for large Stokes numbers than our monofluid implementation, it remains severely limited by the grid resolution, leading to numerical artefacts in  of high dust density.

\subsubsection*{Intermediate summary}
The main results of our analysis of the dust-ratio variations are as follows:
        \begin{itemize}
                \item Our two-fluid implementation with dust as Lagrangian particles leads to dust enrichment that exceeds the maximum value set by the grid resolution used to compute the drag from the gas (see Fig. \ref{Fig:dust_enrichment_bifluid}).
                \item The high-density regions show dust-ratio variations at all sizes with the two-fluid solver  that were not observed in the monofluid results. 
        \end{itemize}

\subsection{Adaptive resolution length PDFs}

We define the adaptive resolution length of the dust-particle distribution  as
\begin{equation}
        h=\left(\frac{m_\mathrm{p}}{\rho_{\mathrm{d},i}}\right)^{1/3},
\end{equation} 
where $\rho_{\mathrm{d},i}$ is the local density associated with each particle, which is computed following the procedure described in Sect.~\ref{Sec:postprocessing}. Therefore, $h$ is a measure of the highest resolution reached in the particle distribution, to be compared with the grid scale $\Delta x$.

\begin{figure}[t!]
        \includegraphics[width=0.5\textwidth]{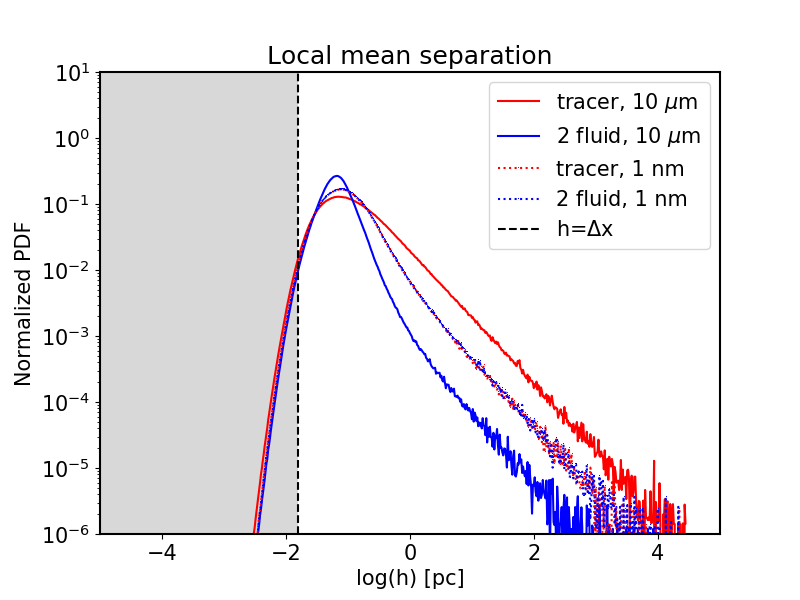}

        \caption{PDF of the effective adaptive resolution length $h$ of the $1$~nm (dotted lines) and  $10~\mu$m (solid lines) dust-grain fluid particle distributions for the 256\_2F run. The colour coding indicates the particle distribution (two-fluid dust as  Lagrangian particles in blue, tracer in red). The vertical line indicates the numerical resolution of the grid ($256^3$). }
        \label{Fig:mean_hdust_pdf}
\end{figure}

Figure \ref{Fig:mean_hdust_pdf} shows the PDFs of the adaptive resolution length $h$ obtained for the particle distributions of run 256\_2F. The scales covered range over more than six orders of magnitude. At large values (high densities), the PDFs of the 10~$\mu$m behave very differently. The two-fluid dust  Lagrangian-particle distribution is more compact, while the particle distribution of the tracer exhibits larger voids. Conversely, all the PDFs show the same trend at small scales, namely high densities. Interestingly, when we compare the adaptive resolution length to the resolution of the grid used for the gas dynamics, we observe that for $h<2\Delta x$, the PDFs are almost identical, independently of the dust-grain size and integration method (two-fluid dust Lagrangian or tracer particles). This indicates that the dust-grain dynamics at scales smaller than the grid size is dominated by numerical artefacts and the clustering is therefore artificial. Indeed, the gas drag is the only mechanism that can move dust particles, and the gas drag cannot be accurately computed on scales smaller than $\Delta x$. As a consequence, the  clusters of high dust density are characteristic of regions dominated by numerical diffusion, which leads to the conclusion that tiny grain clustering observed in the particle distributions has a numerical origin. It remains unclear as to what extent this clustering affects the rest of the simulation volume, in particular the low-density tail of PDF.

\begin{figure*}[t!]
        \includegraphics[width=0.33\textwidth]{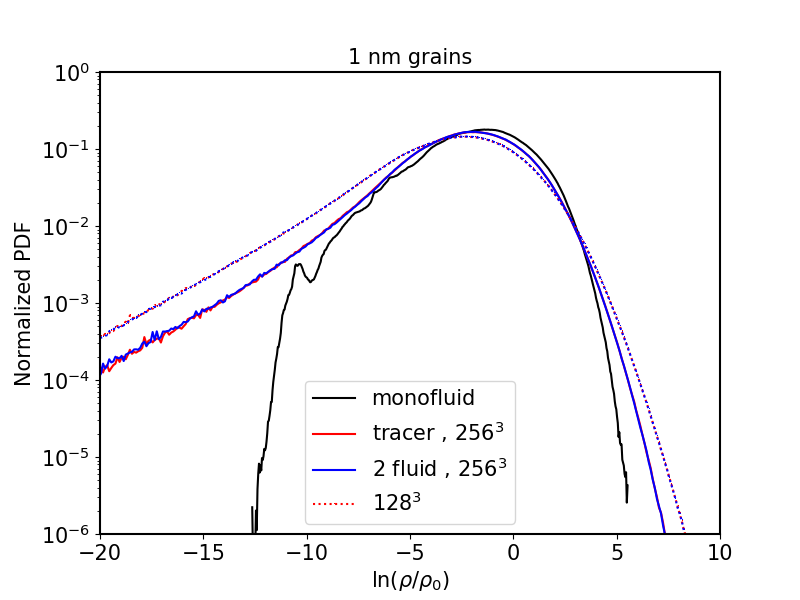}
        \includegraphics[width=0.33\textwidth]{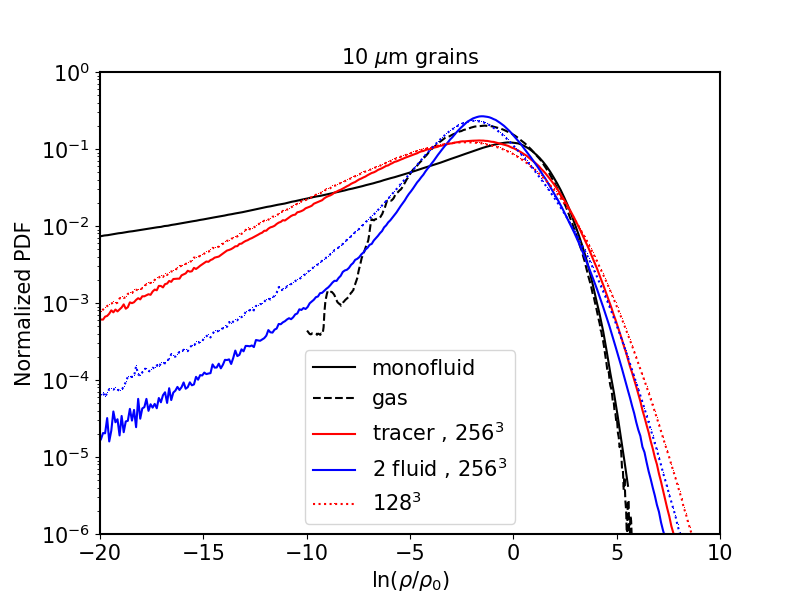}
        \includegraphics[width=0.33\textwidth]{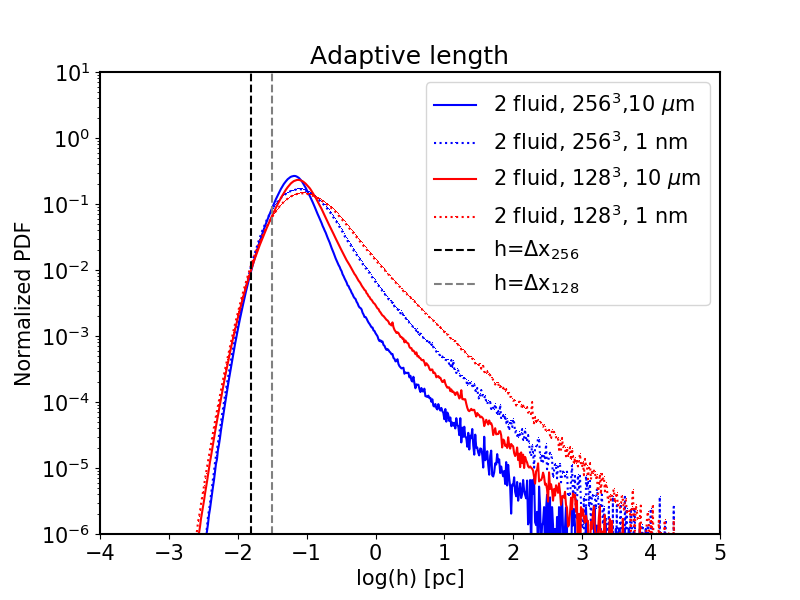}
        \caption{PDF of the dust density of the 1~nm  (left) and $10~\mu$m (middle). The colour coding indicates the dust fluid treatment (two fluid  Lagrangian particles in blue, tracer particles  in red, monofluid in black). For clarity, only the monofluid quantities of the 256\_2F are shown. The solid lines show the results of the 256\_2F run, while the dotted lines show the results of the 128\_2F. {\it Right:} PDF of the adaptive length of the two-fluid dust Lagrangian-particle distributions (solid lines for the $10~\mu$m dust grains, dotted line for the  1~nm ) for the 256\_2F (blue) and 128\_2F (red) models. The vertical dashed lines indicate the grid numerical resolution (black for 256\_2F, grey fro 128\_2F). }

        \label{Fig:res_pdf}
\end{figure*}

\begin{figure*}[t!]
        \includegraphics[width=0.33\textwidth]{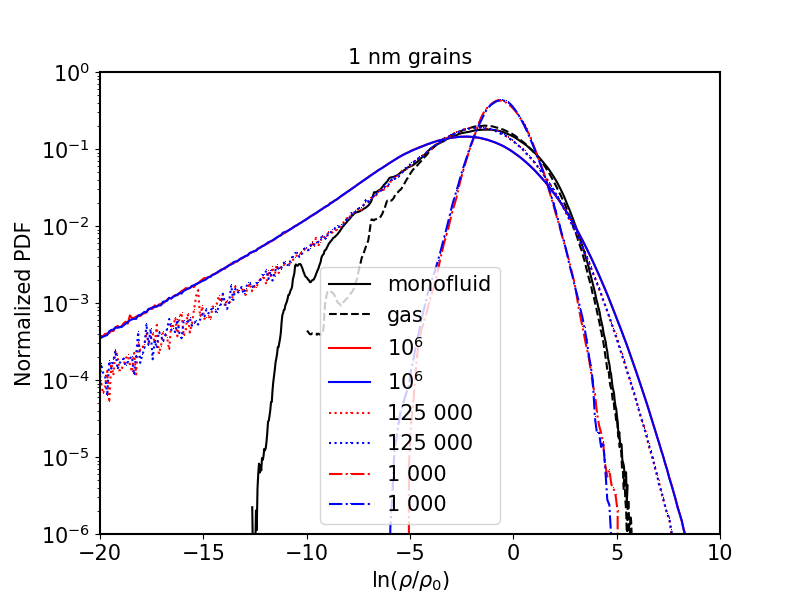}
        \includegraphics[width=0.33\textwidth]{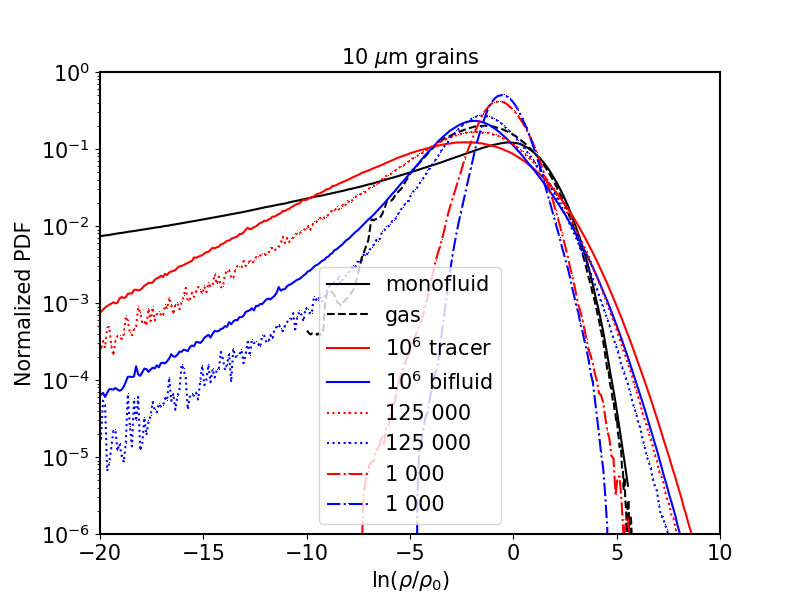}
        \includegraphics[width=0.33\textwidth]{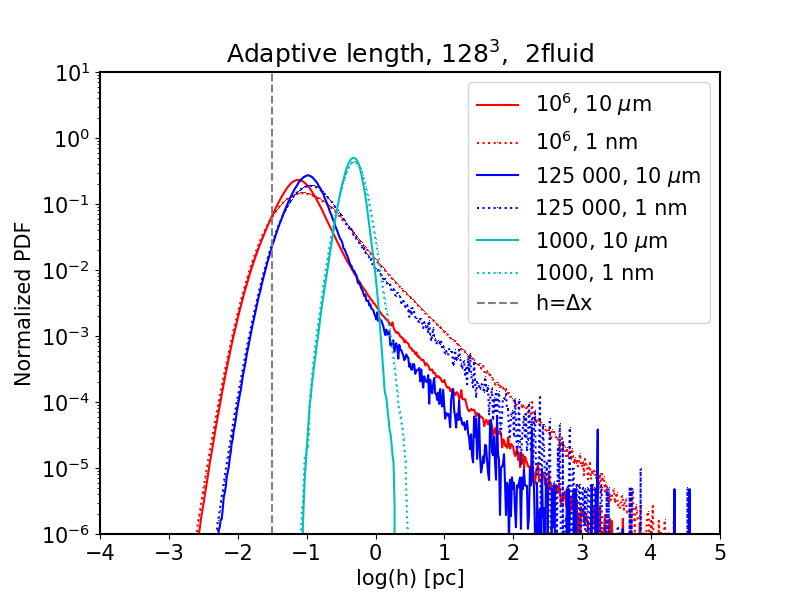}
        
        \caption{Same as Fig. \ref{Fig:res_pdf} for the 128\_2F, 128\_2F\_LR, and 128\_2F\_VLR models. In the density PDFs, the blue lines indicate the results for the two fluid dust  Lagrangian-particle distribution while the red lines indicate the tracer particle ones. }
        \label{Fig:npart_pdf}
\end{figure*}

\subsubsection*{Intermediate summary}
        The main results of the analysis of the Lagrangian-particle clumping scale are as follows: 
        \begin{itemize}
                \item The two-fluid Lagrangian-particle dust results show clustering on scales smaller than the grid size, where the physics of the dust dynamics is not resolved. This behaviour is not physical, but numerical in nature.
                \item All dust sizes show the same trend of artificial clustering below the grid scale.
        \end{itemize}

\subsection{Effects of the grid resolution and the number of particles }

In this section, we investigate the effect of varying the grid resolution $N$  and the number particles $N_\mathrm{p}$ in order to test the effect of the relative resolution used for dust particles in comparison with the relative resolution of the gas. First, we compare the 256\_2F and 128\_2F runs, in which the only difference is the resolution used for the grid. Figure \ref{Fig:res_pdf} shows the PDFs of the 1~nm  and $10~\mu$m density and the adaptive resolution length of the  dust-particle distributions in these two models. For the 1~nm dust grains, the PDFs of the dust density are identical regardless of the treatment (two fluid vs. tracer)  at a given resolution. The PDFs of the 128\_2F are wider compared to the 256\_2F ones, which indicates a more important decoupling from the gas dynamics as the gas resolution decreases. For the $10~\mu$m dust grains, we observe the same trend, but with different PDFs for the two-fluid  Lagrangian-particle distribution and tracer particle  distribution, as already reported in Sect.~\ref{Sec:bifluid_density}. From the adaptive length PDFs, we observe that the 128\_2F exhibits the largest clustering effect. One would have naively expected the 128\_2F model to give better results than the 256\_2F model for the dust-particle distribution, as the particle resolution per grid cell is eight higher in the 128\_2F model. However, as demonstrated in the previous section, the dust-particle dynamics is strongly affected by the numerical resolution used for the gas on the grid, and the amount of artificial clustering is higher.  

Figure \ref{Fig:npart_pdf}  shows the PDFs of the 1~nm  and $10~\mu$m dust-grain density and adaptive resolution length of the  dust-particle distributions in the 128\_2F ($10^6$ particles per grain size bin), 128\_2F\_LR ($125\times10^3$ particles), and 128\_2F\_VLR ($10^3$ particles) models. The aim is to test the effect of the number of particles used to describe the dust fluid at a given resolution for the gas. We see that the dust-density PDFs get tighter as the number of particles decreases. The best match between the two-fluid and the monofluid results is found for the 128\_2F\_LR run for the 1~nm dust grains. For the 128\_2F\_VLR model, the results for the  1~nm  and $10~\mu$m dust grains are qualitatively similar: the resolution is insufficient to correctly capture the dust-grain dynamics. Finally, the minimum adaptive resolution length increases when the number of particles decreases.

Our  short resolution study does not indicate a convergence on the particle-distribution properties. However, it is clear that the properties of the two-fluid  Lagrangian-particles and tracer-particle evolution strongly depend on the grid resolution, as well as on the total number of particles used to sample the dust fluid.  We discuss the limits of our different methods used to follow the dust dynamics in the following section.

\subsubsection*{Intermediate summary}
                The main results of our analysis of the two-fluid Lagrangian-particle dust method convergence are as follows:             
        \begin{itemize}
                \item The two-fluid results are highly sensitive to both the grid resolution and the total number of particles.
                \item Our resolution study of the grid and particle resolutions does not show convergence.
        \end{itemize}

\section{Discussion}

\subsection{Limits of the monofluid and two-fluid formalisms}

Our results  presented in Sect.~\ref{Sec:monofluid} show that the monofluid approximation is well suited for our setup, even for the largest grains, which exhibit large Stokes numbers $\mathrm{St}>1$ in significant parts of the computational volume. For all dust sizes, the regions where $\mathrm{St}>0.1$  show dynamical size sorting and dust-ratio variations of orders of magnitude. However,  the amplitude of the largest dust-ratio variations we report for grains  with sizes $s\geq1~\mu$m  is questionable, because the diffusion approximation we use  is not well suited for large Stokes numbers $\mathrm{St}>1$ .  However, the mass enclosed in these regions remains negligible compared to the total mass ($\simeq 10^{-2}$). In addition, we find no evidence of  systematic effects on the  time evolution of dust enrichment, because the turbulence resets the flow within a crossing time. More importantly, we show that  dust grains with sizes $s\geq4~\mu$m  significantly decouple from the gas, which is consistent with the critical size of  $s\gtrapprox 7~\mu$m that we derive in Sect. \ref{Sec:expected_limits} for our setup and $\mathrm{St}>0.1$. Finally, our monofluid solver does account for the back-reaction of the dust on the gas dynamics. We show in Appendix~\ref{Sec:app_BR} that the back-reaction crucially affects the dust enrichment of the smallest grains (nanometers), which is concentrated in the  regions of low gas density.  Indeed, as large dust grains accumulate, the gas gets pushed away from the 
regions of high dust concentration. The smallest dust grains, which are well coupled to the gas, are therefore expelled. We do not observe this high concentration at low density in our models that neglect the back-reaction. 
Our numerical  experiments and implementation of the monofluid equations using the terminal velocity approximation do not allow us to investigate the dynamics of grains larger than $10~\mu$m. An alternative implementation, such as full monofluid \citep{laibe:14a,laibe:14b}, a full two-fluid Eulerian treatment \citep{benitez:19}, or the two-fluid dust as Lagrangian superparticles we employed (with the addition of dust back-reaction), should better handle high Stokes numbers. We note that for Stokes numbers a few times above unity, the fluid formalism itself becomes questionable.

Using the two-fluid dust as  Lagrangian particles, our results are at odds with the monofluid ones for  all dust sizes. Firstly, for the smallest dust-grain sizes, the bulk and the peak of the monofluid and two-fluid density PDFs are quantitatively comparable. However, the two-fluid PDF tails extend to much higher and lower densities (more than one order of magnitude). The two-fluid particles indeed encounter the same limitation as classical tracer particles, which experience artificial trapping in the high-density regions \citep{cadiou:19}. This leads to an apparent decoupling of the dynamics of the dust particles and the gas, which then appears as large variations in the dust-to-gas ratio. However, this decoupling is of numerical origin, and is mostly due to the finite resolution used for the gas, that is, the grid cell size is larger than the adaptive smoothing length computed from the Lagrangian particle distributions. In addition, we neglect the dust back-reaction in our two-fluid  Lagrangian-particle implementation, and our monofluid results indicate that the smallest grains should not show evidence of large dust enrichment at low gas density. Secondly, for the large grains, the PDF of the density distribution does not converge between the monofluid and  the two-fluid dust as  Lagrangian particles. There is an important discrepancy between the low-density PDF of the two methods. Considering only the two-fluid Lagrangian-particle distribution, at high density, the PDF of the largest grains matches that of the smallest grains, with the same issue of $h<\Delta x$.  We also defined a maximum dust enrichment as a function of the gas density, which is a function of the grid size. From   Fig.~\ref{Fig:dust_enrichment_bifluid}, it is clear that the regions of  maximum dust enrichment, that is, the regions with the highest dust density, are regulated by the grid size $\Delta x$, in particular for the largest grains.

\cite{laibe:12a,laibe:14a} showed that two-fluid methods are limited by both temporal and spatial resolution in the strong drag regime. Firstly, the spatial resolution should verify the criterion $\Delta x < c_\mathrm{s}t_\mathrm{s}$ in order to accurately capture the dust and gas dynamical phase separation, and to avoid artificial velocity difference damping. It is straightforward to estimate the dust-grain size $s_\mathrm{bifluid}$ which satisfies $\Delta x < c_\mathrm{s}t_\mathrm{s}$ using (\ref{eq:ts}):
\begin{equation}
        s_\mathrm{bifluid} > \Delta x\frac{\rho_\mathrm{g}}{\rho_\mathrm{grain}} \sqrt{\frac{8}{\pi \gamma}},
\end{equation}
or, in typical conditions of molecular clouds,
\begin{equation}
        s_\mathrm{bifluid} > 380~\mu\mathrm{m} \left( \frac{\Delta x}{1~\mathrm{pc}}\right)\left(\frac{\rho}{10^{-20}~\rm{g}~\rm{cm}^{-3}}\right)\left( \frac{\rho_\mathrm{grain}}{1~\rm{g}~\rm{cm}^{-3}}\right)^{-1},
        \label{eq:sbifluid}
\end{equation}
where $\Delta x$ is the characteristic resolution length of the gas fluid.  This criterion might have been overlooked in previous studies. In our experiments, we neglected the dust back-reaction, meaning that the over-damping is not problematic \citep{loren:15}. We verified this by running some \ttfamily{DUSTYWAVE }\rm experiments  \citep{laibe:12a} with our monofluid and two-fluid solvers with a tiny ($\epsilon=10^{-5}$) dust fraction. In the strong drag regime, we find that both methods give accurate results.  In addition to the spatial resolution criterion, \cite{laibe:12a,laibe:14a} showed that the integration time-step constraint due the very short stopping time conditions can be very restrictive. In this work, we solve this issue by using an analytical update of the dust particle velocity. In addition, the characteristic velocity of the gas in supersonic turbulence experiments  is larger than the sound speed, meaning that the condition in Eq. (\ref{eq:sbifluid}) relaxes towards smaller values, because $\Delta x < v_\rms t_\mathrm{s}$. Assuming Larson scaling relations, we get
\begin{equation}
        s_\mathrm{bifluid} > 84~\mu\mathrm{m} \left( \frac{\Delta x}{1~\mathrm{pc}}\right)\left(\frac{\rho}{10^{-20}~\rm{g}~\rm{cm}^{-3}}\right)^{12/7}\left( \frac{\rho_\mathrm{grain}}{1~\rm{g}~\rm{cm}^{-3}}\right)^{-1}.
\end{equation}

For the two-fluid results, it is  important to note that we have found no systematic accumulation of the particles in the  regions of high gas density in the two-fluid  Lagrangian-particle population. Indeed, the gas flow, being highly turbulent, changes drastically over one crossing time, such that the artificially trapped particles can be released in the low-density material. A hybrid treatment of the particle motion in the high-density regions could help to alleviate this problem by superimposing the dust-mass fluxes  from the Eulerian  description at small Stokes numbers onto the Lagrangian model \citep[see e.g.][]{cadiou:19}. 

From the study we conduct here, with its  limited resolution, we observe a strong dependence of the dust density PDFs and clustering degree on the number of particles and grid resolution. Indeed, the number of particles as well as the number of grid cells should increase, as should the ratio between the two, similarly to what is expected in SPH codes with the number of particles and the number of neighbours \citep[e.g.][]{commercon:08}.

\subsection{Comparison to previous works}

Using SPH, \cite{tricco:17} employed the same approximations as ours, that is, monofluid and diffusion approximation, and used very similar initial conditions and numerical resolution. In the context of compressible turbulence, it is worth remembering that \cite{price:10} report that SPH and grid codes give roughly comparable results when the number of SPH particles is approximately equal to the number of grid cells.
We can therefore directly compare our results with those of \cite{tricco:17}.  The models of these latter authors use a grain density of 3 g cm$^{-3}$, while we use 1 g cm$^{-3}$. As a consequence, at equal Stokes number, our dust size has to be multiplied by a factor of three in order to compare with their results. 
Our results qualitatively agree with those of \cite{tricco:17}, who report that grains with size $>10~\mu$m decouple significantly from the gas dynamics. We find that the dust grains with size $>4~\mu$m decouple significantly (which thus corresponds to $12~\mu$m in the experiments of  \cite{tricco:17}), with large variations in the dust concentration (more than one order of magnitude). In terms of initial Stokes number, this corresponds to $\mathrm{St}\gtrsim 0.1$. This behaviour is referred to as size-sorting in \cite{tricco:17}. We find that the size-sorting most efficiently occurs in  regions of intermediate gas density ($-4<\log(\rho_\mathrm{g}/\rho_\mathrm{g,0})<1$), whereas the densest parts of the molecular clouds exhibit uniform dust concentration, equal to the initial dust concentration. For the small grains, with sizes $<4~\mu$m, we report some variation in the local dust ratio, but these regions represent a negligible amount of mass and correspond to gas densities where $\rho_\mathrm{g}/\rho_\mathrm{g,0}<1$.
We can therefore expect apparent variation of the dust-size distribution in these intermediate densities. However,  when observed in real conditions with an integration along the line of sight, this inhomogeneity will be smoothed out by the bulk of the mass, which has a uniform dust concentration. 

In order to compare our results with those of  \cite{hopkins:16} and \cite{mattsson:19a,mattsson:19b}, we introduce the dimensionless parameter $\alpha$ as defined in \cite{hopkins:16}:
\begin{equation}
        \alpha=\frac{\rho_{\mathrm{grain}}s}{<\rho_\mathrm{gas}> L},
\end{equation}
where $<\rho_\mathrm{gas}>\simeq \rho_0$ is the mean density of the gas. In our numerical setup, this translates to 

\begin{equation}
        \alpha\simeq 0.002 \left(\frac{s}{1~\mu\mathrm{m}}\right) \simeq 0.12 ~\mathrm{St}.
\end{equation}

\cite{hopkins:16} and \cite{mattsson:19a,mattsson:19b} report  a small-scale clustering reaching its maximum for grains in the nanometer range (a few tens of nanometers), which corresponds to $\alpha\simeq0.01 -0.1$ for their parameter choice.
In our setup, we  investigate a parameter space for which $\alpha<0.02$, corresponding to the  very small grains in \cite{hopkins:16} and \cite{mattsson:19a,mattsson:19b}.

\cite{hopkins:16} designed their numerical experiment to investigate gas and dust dynamics in giant molecular clouds. These authors used a $256^3$ particle resolution for both the dust and the gas fluids, using the mesh-free \ttfamily{GIZMO }\rm  framework \citep{hopkins:15}. The numerical setup used by \cite{mattsson:19a,mattsson:19b}  is more similar to ours (gas dynamics on a fixed grid, dust treated as Lagragian superparticles), but they  investigate dust grains with $\alpha>0.1$. Our results for the two-fluid dust as  Lagrangian particles are globally consistent with those of \cite{hopkins:16} at very low $\alpha$, with the largest variation of the dust ratio observed in the  regions of low gas density. In our two-fluid models, the maximum dust ratio is set by the grid resolution used for the gas dynamics. In the
regions of low gas density, we report a dust ratio increase of up to four orders of magnitude for the $\alpha\simeq 0.02$ dust grains, while \cite{hopkins:16} found variations of up to six orders of magnitude for $\alpha= 0.03$. \cite{hopkins:16} used a completely different numerical framework from ours, with adaptive resolutions of both the gas and the dust particles, meaning that they can better manage different resolutions for the gas and the dust (despite still being limited by the gas resolution for the drag). Further work at higher resolution is needed in order to investigate the behaviour of the $\alpha\simeq 0.02$ dust grains in the  material of low gas density with our setup. 
At high gas density, we observe dust-ratio variations over more that two orders of magnitude, also similar to \cite{hopkins:16} results. We attribute these variations to spurious effects due to  trapping of Lagrangian superparticles beyond the grid scale, which gives rise to an artificial increase in the dust-fluid density. Our results therefore question these fluctuations at high density, which are typically regions where $\mathrm{St}<1$. These correspond to  adaptive resolution lengths of the Lagrangian-particle distributions smaller that the grid size  and where the grid size is larger than the coupling length. The effects of these spurious variations at small Stokes number on the dust ratio at low density remains unclear from our comparison.  \cite{hopkins:14} and \cite{hopkins:16} suggest that the variations of stellar abundances within clusters could be a consequence of the dust dynamical coupling at the  scales of giant molecular cloud.  In particular, \cite{hopkins:14} proposes that decoupling  due to dust dynamics could lead to the formation of totally metal stars (1 for $10^4$ in clusters). Our two-fluid results show that the maximum dust enrichment for the largest dust  grains is regulated by the  size of the grid used to  interpolate the friction with the gas. On the other hand, our monofluid results show that the maximum dust enrichment for the largest grain corresponds to regions where $\mathrm{St}>1$, that is, regions where the TVA approximation is out of range. In summary, our results do not allow us to make a firm estimation of the amplitude of dust-ratio variations for dust grains with $\mathrm{St}>1$ .  Greater resolution is still clearly needed in order to quantify the occurrence of metal stars as suggested by \cite{hopkins:14}. Two-fluid Eulerian dust simulations should also be used as the method does not cause artificial clustering for particles with $\mathrm{St}\gtrsim1$.

More recently, \cite{mattsson:19b} performed similar experiments to ours, using a $1024^3$ grid resolution for the gas and $10^6-10^7$ particles for each dust-size bin. These authors compare the properties of the  two-fluid Lagrangian superparticles with the those of the tracer particles for the smallest grains they consider ($\alpha \simeq 0.001$) and find that the inertia still makes a difference in the  grain dynamics and small-scale clustering. Our comparison between the tracer particles and two-fluid Lagrangian superparticles shows that this is not the case for very small Stokes number, as both particle distributions are identical. More importantly, we show that  independently of the dust size, the regions where $\mathrm{St}<1$ show evidence of artificial dust-particle trapping (or clustering) similar to that of the tracer particles. Our findings therefore bring into question the small-scale clustering properties of nano-dust grains reported by  \cite{mattsson:19b}, in particular with respect to the minimum resolution set by the grid. 

This short comparison points towards the need for a hybrid numerical method for dust-particle mesh codes in order to handle the large variety of Stokes numbers encountered locally in turbulent molecular clouds. For dust-grain species with both $\mathrm{St}>1$ in the material with low gas density and $\mathrm{St}<1$ in that with high gas density, one could use our two-fluid implementation for $\mathrm{St}>1$ to get the particle acceleration, but a Monte Carlo scheme based on the Godunov flux at cell interface for  $\mathrm{St}<1$ as proposed by  \cite{cadiou:19}.

\subsection{Neglected physics}

In this work, we studied the dynamical interplay between dust grains and gas in conditions typical of molecular clouds. We made  a number of simplifications in order to focus on numerical implementation effects rather than on physical effects. The first big assumption we made concerns magnetic fields. Magnetic fields are ubiquitously observed in molecular clouds with micro-Gauss amplitudes \citep{hennebelle:12,pattle:19} and can interplay with the turbulent motions  of the gas \citep{federrath:16}. In our setup, accounting for magnetic fields would have two implications. First, the properties of the  turbulent flows change in the presence of magnetic fields, in particular in the strong field case, with anisotropies arising from the strong magnetic tension. However, this should not alter our findings as to the reliability of the monofluid and two-fluid formalisms, as long as dust grains are considered to be neutral. \cite{hopkins:16} report that including magnetic fields has a weak effect on the amplitudes of the gas-to-dust-ratio variations, but that  the dust density can vary in regions where the gas density does not vary. Indeed, the neutral dust grains are sensitive to the gas-pressure gradients and Lorentz force of the gas through the drag. In shearing-box experiments of dust dynamics within protoplanetary discs, \cite{lebreuilly:23} show that dust grains preferentially concentrate in regions where the pressure-gradient force and Lorentz force cancel out. Second, dust grains are expected to be charged in molecular clouds \citep[e.g.][]{draine:87,guillet:07}, and depending on the dust-grain size, either the Lorentz force or the gas drag should regulate the dust-grain motions. The balance between the gyration time and the stopping time of the dust grains of different sizes can therefore favour dust-size sorting. \cite{lee:17} extended the work of \cite{hopkins:16} to include the dynamics of charged dust grains and find that the Lorentz force suppresses the dust-ratio variations for very small grains ($s\ll 1 ~\mu$m), while the large grains ($s\gtrsim 1 ~\mu$m) are insensitive to magnetic fields. These authors concluded that  the effects of Lorentz forces are subdominant, and so the qualitative conclusions from their previous studies remain unchanged. However, they expect that the dynamics of charged dust grains would be more important in physical conditions other than the cold neutral medium, such as in the warm neutral medium. Exploration of the dynamics of charged dust grains with the monofluid formalism still lacks a proper derivation of the monofluid equation accounting for the Lorentz acceleration and the dust back-reaction. 

The second strong assumption is on dust coagulation and fragmentation processes. We limit our physical model to dust dynamical evolution, in which only the interactions between the dust and the gas are taken into account. The supersonic motions within molecular clouds associated with the dust size sorting  favour strong interactions between dust grains, in particular collisions that can lead to dust growth and fragmentation. However, we show that for dust-grain sizes typically observed in molecular clouds \citep[$<1~\mu$m,][]{kohler:15}, only a small mass fraction of the dust grains  decouples from the gas and could therefore be undergoing grain--grain interactions. Further work should account for dust-grain interaction in order to test whether or not dust growth can already occur in molecular clouds thanks to the dynamical size sorting as suggested by previous work by \cite{ormel:09}, who report growth up to $100~\mu$m in clouds with lifetimes larger than a freefall time.

\section{Conclusion}

We present a unique suite of numerical experiments designed to investigate neutral dust-grain dynamics in turbulent molecular clouds. We considered typical turbulence driving parameters, which satisfy the classical Larson relations \citep{larson:81,hennebelle:12}. We compared two different numerical implementations of the dust dynamics. The first one relies on the monofluid formalism and the diffusion approximation. The second implementation treats the dust fluid using Lagrangian super-particles, while the gas dynamics is handled on a Eulerian grid. For each implementation, we compared the dust-density distributions and the spatial dust-ratio variations as a function of dust-grain size. 
Our main findings are as follows:

\begin{itemize}
\item
Our implementation of the monofluid formalism, which relies on the dust diffusion approximation \citep{price:15,lebreuilly:19}, is well suited to studying the dynamics of neutral dust grains with sizes $1~\mathrm{nm}<s<10~\mu$m. We report dust dynamics decoupling for Stokes numbers $\rm{St}>0.1$, that is, dust grains of $s>4~\mu$m in size, which matches the theoretical expectations provided in Eq.~\ref{eq:scrit}. Our results are in very good agreement with previous work by \cite{tricco:17}. The dust concentrates in the pressure maxima, with a higher concentration at low density $(\rho<\rho_0)$.\\
\item The  two-fluid results with dust as Lagrangian particles are affected by numerical artefacts  in both the strong and weak drag regimes as follows. For tightly coupled dust grains, with $\rm{St}\ll1$, we  show that the apparent large dust-to-gas ratio variations are spurious, mostly because of artificial trapping in the high-density regions. For weakly coupled dust grains in the $\rm{St}>1$ regime, the maximum dust enrichment we measure is strongly affected by the grid resolution, from which the drag from the gas is interpolated. These results raise questions about the robustness of dust concentration and clustering predictions measured in numerical experiments of dusty turbulence using a hybrid multifluid numerical method (gas on a grid, dust as Lagrangian particles). Our resolution study shows that  results from a setup with dust as Lagrangian particles do not converge to the correct results with increasing grid and particle resolutions. Instead of deriving resolution criteria, a more fundamental study of the correctness of the Lagrangian implementation at different drag regimes should be carried out.\\
\item We discuss the comparison  with previous works and we show that there is no tension in terms of the critical size for decoupling between the results reported by \cite{tricco:17} on one side and \cite{hopkins:16} and \cite{mattsson:19a} on the other. These studies did not explore the same parameter space as suggested in footnote 13 of \cite{hopkins:20}. The controversy arose from trying to conclude on a universal dust size for decoupling, independently of the physical conditions. Our results show that these previous studies are in good agreement when the comparison is done at equivalent Stokes number or $\alpha$ parameter.\\
\item If one takes typical dust grains in molecular clouds to have sizes in agreement with the standard MRN size distribution, namely a maximum dust size of $<1~\mu$m, then our results indicate that the dust-to-gas ratio should not exhibit large variations of more than a factor of two  in  regions of high gas density. We stress that these results are valid for (1) neutral dust grains and (2) turbulent properties following the Larson relations at parsec scales. 
\end{itemize}

\begin{acknowledgements}
 This work was granted access to the HPC resources of CINES (Occigen) under the allocation 2018-047247 made by GENCI. We gratefully acknowledge support from the PSMN (Pôle Scientifique de Modélisation Numérique) of the ENS de Lyon.  BC and FL have received fundings from  the French national agency ANR DISKBUILD  ANR-20-CE49-0006. This work was supported with funding from the European Union’s Horizon 2020 research and innovation programme under the Marie Sklodowska Curie grant agreement No 823823 (RISE DUSTBUSTERS project). BC acknowledges financial support from "Programme National de Physique Stellaire" (PNPS) of CNRS/INSU, CEA and CNES, France. UL acknowledges financial support from the European Research Council (ERC) via the ERC Synergy Grant ECOGAL (grant 855130).
All the figures were created using the \href{https://github.com/osyris-project/osyris}{\texttt{OSYRIS}} visualisation package for \texttt{RAMSES}.
\end{acknowledgements}

\bibliographystyle{aa}
\bibliography{biblio}

\begin{thebibliography}{64}
\expandafter\ifx\csname natexlab\endcsname\relax\def\natexlab#1{#1}\fi

\bibitem[{{Bate} \& {Lor{\'e}n-Aguilar}(2017)}]{bate:17}
{Bate}, M.~R. \& {Lor{\'e}n-Aguilar}, P. 2017, \mnras, 465, 1089

\bibitem[{{Ben{\'\i}tez-Llambay} {et~al.}(2019){Ben{\'\i}tez-Llambay}, {Krapp},
  \& {Pessah}}]{benitez:19}
{Ben{\'\i}tez-Llambay}, P., {Krapp}, L., \& {Pessah}, M.~E. 2019, \apjs, 241,
  25

\bibitem[{{Cadiou} {et~al.}(2019){Cadiou}, {Dubois}, \& {Pichon}}]{cadiou:19}
{Cadiou}, C., {Dubois}, Y., \& {Pichon}, C. 2019, \aap, 621, A96

\bibitem[{{Commer{\c{c}}on} {et~al.}(2008){Commer{\c{c}}on}, {Hennebelle},
  {Audit}, {Chabrier}, \& {Teyssier}}]{commercon:08}
{Commer{\c{c}}on}, B., {Hennebelle}, P., {Audit}, E., {Chabrier}, G., \&
  {Teyssier}, R. 2008, \aap, 482, 371

\bibitem[{{Commer{\c{c}}on} {et~al.}(2019){Commer{\c{c}}on}, {Marcowith}, \&
  {Dubois}}]{commercon:19}
{Commer{\c{c}}on}, B., {Marcowith}, A., \& {Dubois}, Y. 2019, \aap, 622, A143

\bibitem[{{Coutens} {et~al.}(2020){Coutens}, {Commer{\c{c}}on}, \&
  {Wakelam}}]{coutens:20}
{Coutens}, A., {Commer{\c{c}}on}, B., \& {Wakelam}, V. 2020, \aap, 643, A108

\bibitem[{{Dipierro} {et~al.}(2015){Dipierro}, {Price}, {Laibe}, {Hirsh},
  {Cerioli}, \& {Lodato}}]{dipierro:15}
{Dipierro}, G., {Price}, D., {Laibe}, G., {et~al.} 2015, \mnras, 453, L73

\bibitem[{Draine(2003)}]{draine:03}
Draine, B. 2003, Annual Review of Astronomy and Astrophysics, 41, 241

\bibitem[{{Draine} \& {Salpeter}(1979)}]{draine:79}
{Draine}, B.~T. \& {Salpeter}, E.~E. 1979, \apj, 231, 77

\bibitem[{Draine \& Sutin(1987)}]{draine:87}
Draine, B.~T. \& Sutin, B. 1987, ApJ, 803

\bibitem[{{Epstein}(1924)}]{epstein:24}
{Epstein}, P.~S. 1924, Physical Review, 23, 710

\bibitem[{Eswaran \& Pope(1988)}]{eswaran:88}
Eswaran, V. \& Pope, S.~B. 1988, Computers {\&} Fluids, 16, 257

\bibitem[{{Federrath}(2016)}]{federrath:16}
{Federrath}, C. 2016, Journal of Plasma Physics, 82, 535820601

\bibitem[{{Federrath} {et~al.}(2010){Federrath}, {Roman-Duval}, {Klessen},
  {Schmidt}, \& {Mac Low}}]{federrath:10}
{Federrath}, C., {Roman-Duval}, J., {Klessen}, R.~S., {Schmidt}, W., \& {Mac
  Low}, M.-M. 2010, \aap, 512, A81

\bibitem[{Galametz {et~al.}(2019)Galametz, Maury, Valdivia, Testi, Belloche, \&
  Andr{\'{e}}}]{galametz:19}
Galametz, M., Maury, A.~J., Valdivia, V., {et~al.} 2019, Astronomy and
  Astrophysics, 632 [\eprint[arXiv]{1910.04652}]

\bibitem[{{Guillet} {et~al.}(2018){Guillet}, {Fanciullo}, {Verstraete},
  {Boulanger}, {Jones}, {Miville-Desch{\^e}nes}, {Ysard}, {Levrier}, \&
  {Alves}}]{guillet:18}
{Guillet}, V., {Fanciullo}, L., {Verstraete}, L., {et~al.} 2018, \aap, 610, A16

\bibitem[{{Guillet} {et~al.}(2020){Guillet}, {Hennebelle}, {Pineau des
  For{\^e}ts}, {Marcowith}, {Commer{\c{c}}on}, \& {Marchand}}]{guillet:20}
{Guillet}, V., {Hennebelle}, P., {Pineau des For{\^e}ts}, G., {et~al.} 2020,
  \aap, 643, A17

\bibitem[{{Guillet} {et~al.}(2007){Guillet}, {Pineau Des For{\^e}ts}, \&
  {Jones}}]{guillet:07}
{Guillet}, V., {Pineau Des For{\^e}ts}, G., \& {Jones}, A.~P. 2007, \aap, 476,
  263

\bibitem[{{Hennebelle} \& {Falgarone}(2012)}]{hennebelle:12}
{Hennebelle}, P. \& {Falgarone}, E. 2012, \aapr, 20, 55

\bibitem[{{Heyer} \& {Brunt}(2004)}]{heyer:04}
{Heyer}, M.~H. \& {Brunt}, C.~M. 2004, \apjl, 615, L45

\bibitem[{{Hopkins}(2014)}]{hopkins:14}
{Hopkins}, P.~F. 2014, \apj, 797, 59

\bibitem[{{Hopkins}(2015)}]{hopkins:15}
{Hopkins}, P.~F. 2015, \mnras, 450, 53

\bibitem[{{Hopkins} \& {Lee}(2016)}]{hopkins:16}
{Hopkins}, P.~F. \& {Lee}, H. 2016, \mnras, 456, 4174

\bibitem[{{Hopkins} {et~al.}(2020){Hopkins}, {Squire}, \&
  {Seligman}}]{hopkins:20}
{Hopkins}, P.~F., {Squire}, J., \& {Seligman}, D. 2020, \mnras, 496, 2123

\bibitem[{{Hutchison} {et~al.}(2018){Hutchison}, {Price}, \&
  {Laibe}}]{hutchison:18}
{Hutchison}, M., {Price}, D.~J., \& {Laibe}, G. 2018, \mnras, 476, 2186

\bibitem[{{K{\"o}hler} {et~al.}(2015){K{\"o}hler}, {Ysard}, \&
  {Jones}}]{kohler:15}
{K{\"o}hler}, M., {Ysard}, N., \& {Jones}, A.~P. 2015, \aap, 579, A15

\bibitem[{{Kwok}(1975)}]{kwok:75}
{Kwok}, S. 1975, \apj, 198, 583

\bibitem[{{Laibe} \& {Price}(2012)}]{laibe:12a}
{Laibe}, G. \& {Price}, D.~J. 2012, \mnras, 420, 2345

\bibitem[{{Laibe} \& {Price}(2014{\natexlab{a}})}]{laibe:14b}
{Laibe}, G. \& {Price}, D.~J. 2014{\natexlab{a}}, \mnras, 444, 1940

\bibitem[{{Laibe} \& {Price}(2014{\natexlab{b}})}]{laibe:14a}
{Laibe}, G. \& {Price}, D.~J. 2014{\natexlab{b}}, \mnras, 440, 2136

\bibitem[{{Larson}(1981)}]{larson:81}
{Larson}, R.~B. 1981, \mnras, 194, 809

\bibitem[{{Lebreuilly} {et~al.}(2019){Lebreuilly}, {Commer{\c{c}}on}, \&
  {Laibe}}]{lebreuilly:19}
{Lebreuilly}, U., {Commer{\c{c}}on}, B., \& {Laibe}, G. 2019, \aap, 626, A96

\bibitem[{{Lebreuilly} {et~al.}(2020){Lebreuilly}, {Commer{\c{c}}on}, \&
  {Laibe}}]{lebreuilly:20}
{Lebreuilly}, U., {Commer{\c{c}}on}, B., \& {Laibe}, G. 2020, \aap, 641, A112

\bibitem[{{Lebreuilly} {et~al.}(2022){Lebreuilly}, {Mac Low},
  {Commer{\c{c}}on}, \& {Ebel}}]{lebreuilly:23}
{Lebreuilly}, U., {Mac Low}, M.~M., {Commer{\c{c}}on}, B., \& {Ebel}, D.~S.
  2022, submitted to \aap

\bibitem[{{Lee} {et~al.}(2017){Lee}, {Hopkins}, \& {Squire}}]{lee:17}
{Lee}, H., {Hopkins}, P.~F., \& {Squire}, J. 2017, \mnras, 469, 3532

\bibitem[{{Lesur} {et~al.}(2022){Lesur}, {Ercolano}, {Flock}, {Lin}, {Yang},
  {Barranco}, {Benitez-Llambay}, {Goodman}, {Johansen}, {Klahr}, {Laibe},
  {Lyra}, {Marcus}, {Nelson}, {Squire}, {Simon}, {Turner}, {Umurhan}, \&
  {Youdin}}]{lesur:22}
{Lesur}, G., {Ercolano}, B., {Flock}, M., {et~al.} 2022, arXiv e-prints,
  arXiv:2203.09821

\bibitem[{Lor{\'{e}}n-Aguilar \& Bate(2015)}]{loren:15}
Lor{\'{e}}n-Aguilar, P. \& Bate, M.~R. 2015, Monthly Notices of the Royal
  Astronomical Society, 454, 4114

\bibitem[{{Love} {et~al.}(1994){Love}, {Joswiak}, \& {Brownlee}}]{love:94}
{Love}, S.~G., {Joswiak}, D.~J., \& {Brownlee}, D.~E. 1994, \icarus, 111, 227

\bibitem[{{Marchand} {et~al.}(2021){Marchand}, {Guillet}, {Lebreuilly}, \& {Mac
  Low}}]{marchand:21}
{Marchand}, P., {Guillet}, V., {Lebreuilly}, U., \& {Mac Low}, M.~M. 2021,
  \aap, 649, A50

\bibitem[{{Mattsson} {et~al.}(2019{\natexlab{a}}){Mattsson}, {Bhatnagar},
  {Gent}, \& {Villarroel}}]{mattsson:19a}
{Mattsson}, L., {Bhatnagar}, A., {Gent}, F.~A., \& {Villarroel}, B.
  2019{\natexlab{a}}, \mnras, 483, 5623

\bibitem[{{Mattsson} {et~al.}(2019{\natexlab{b}}){Mattsson}, {Fynbo}, \&
  {Villarroel}}]{mattsson:19b}
{Mattsson}, L., {Fynbo}, J.~P.~U., \& {Villarroel}, B. 2019{\natexlab{b}},
  \mnras, 490, 5788

\bibitem[{{Muro-Arena} {et~al.}(2018){Muro-Arena}, {Dominik}, {Waters}, {Min},
  {Klarmann}, {Ginski}, {Isella}, {Benisty}, {Pohl}, {Garufi}, {Hagelberg},
  {Langlois}, {Menard}, {Pinte}, {Sezestre}, {van der Plas}, {Villenave},
  {Delboulb{\'e}}, {Magnard}, {M{\"o}ller-Nilsson}, {Pragt}, {Rabou}, \&
  {Roelfsema}}]{muro:18}
{Muro-Arena}, G.~A., {Dominik}, C., {Waters}, L.~B.~F.~M., {et~al.} 2018, \aap,
  614, A24

\bibitem[{Ormel {et~al.}(2011)Ormel, Min, Tielens, Dominik, \&
  Paszun}]{ormel:11}
Ormel, C., Min, M., Tielens, A., Dominik, C., \& Paszun, D. 2011,
  $\backslash$Aap, 532, A43

\bibitem[{{Ormel} {et~al.}(2009){Ormel}, {Paszun}, {Dominik}, \&
  {Tielens}}]{ormel:09}
{Ormel}, C.~W., {Paszun}, D., {Dominik}, C., \& {Tielens}, A.~G.~G.~M. 2009,
  \aap, 502, 845

\bibitem[{{Pattle} \& {Fissel}(2019)}]{pattle:19}
{Pattle}, K. \& {Fissel}, L. 2019, Frontiers in Astronomy and Space Sciences,
  6, 15

\bibitem[{{Pencil Code Collaboration} {et~al.}(2021){Pencil Code
  Collaboration}, {Brandenburg}, {Johansen}, {Bourdin}, {Dobler}, {Lyra},
  {Rheinhardt}, {Bingert}, {Haugen}, {Mee}, {Gent}, {Babkovskaia}, {Yang},
  {Heinemann}, {Dintrans}, {Mitra}, {Candelaresi}, {Warnecke},
  {K{\"a}pyl{\"a}}, {Schreiber}, {Chatterjee}, {K{\"a}pyl{\"a}}, {Li},
  {Kr{\"u}ger}, {Aarnes}, {Sarson}, {Oishi}, {Schober}, {Plasson}, {Sandin},
  {Karchniwy}, {Rodrigues}, {Hubbard}, {Guerrero}, {Snodin}, {Losada},
  {Pekkil{\"a}}, \& {Qian}}]{pencil:21}
{Pencil Code Collaboration}, {Brandenburg}, A., {Johansen}, A., {et~al.} 2021,
  The Journal of Open Source Software, 6, 2807

\bibitem[{{Planck Collaboration} {et~al.}(2011){Planck Collaboration},
  {Abergel}, {Ade}, {Aghanim}, {Arnaud}, {Ashdown}, {Aumont}, {Baccigalupi},
  {Balbi}, {Banday}, {Barreiro}, {Bartlett}, {Battaner}, {Benabed},
  {Beno{\^\i}t}, {Bernard}, {Bersanelli}, {Bhatia}, {Blagrave}, {Bock},
  {Bonaldi}, {Bond}, {Borrill}, {Bouchet}, {Boulanger}, {Bucher}, {Burigana},
  {Cabella}, {Cantalupo}, {Cardoso}, {Catalano}, {Cay{\'o}n}, {Challinor},
  {Chamballu}, {Chiang}, {Chiang}, {Christensen}, {Clements}, {Colombi},
  {Couchot}, {Coulais}, {Crill}, {Cuttaia}, {Danese}, {Davies}, {Davis}, {de
  Bernardis}, {de Gasperis}, {de Rosa}, {de Zotti}, {Delabrouille}, {Delouis},
  {D{\'e}sert}, {Dickinson}, {Donzelli}, {Dor{\'e}}, {D{\"o}rl}, {Douspis},
  {Dupac}, {Efstathiou}, {En{\ss}lin}, {Eriksen}, {Finelli}, {Forni},
  {Frailis}, {Franceschi}, {Galeotta}, {Ganga}, {Giard}, {Giardino},
  {Giraud-H{\'e}raud}, {Gonz{\'a}lez-Nuevo}, {G{\'o}rski}, {Gratton},
  {Gregorio}, {Gruppuso}, {Hansen}, {Harrison}, {Helou}, {Henrot-Versill{\'e}},
  {Herranz}, {Hildebrandt}, {Hivon}, {Hobson}, {Holmes}, {Hovest}, {Hoyland},
  {Huffenberger}, {Jaffe}, {Joncas}, {Jones}, {Jones}, {Juvela},
  {Keih{\"a}nen}, {Keskitalo}, {Kisner}, {Kneissl}, {Knox}, {Kurki-Suonio},
  {Lagache}, {Lamarre}, {Lasenby}, {Laureijs}, {Lawrence}, {Leach}, {Leonardi},
  {Leroy}, {Linden-V{\o}rnle}, {Lockman}, {L{\'o}pez-Caniego}, {Lubin},
  {Mac{\'\i}as-P{\'e}rez}, {MacTavish}, {Maffei}, {Maino}, {Mandolesi}, {Mann},
  {Maris}, {Marshall}, {Martin}, {Mart{\'\i}nez-Gonz{\'a}lez}, {Masi},
  {Matarrese}, {Matthai}, {Mazzotta}, {McGehee}, {Meinhold}, {Melchiorri},
  {Mendes}, {Mennella}, {Miville-Desch{\^e}nes}, {Moneti}, {Montier},
  {Morgante}, {Mortlock}, {Munshi}, {Murphy}, {Naselsky}, {Nati}, {Natoli},
  {Netterfield}, {N{\o}rgaard-Nielsen}, {Noviello}, {Novikov}, {Novikov},
  {O'Dwyer}, {Osborne}, {Pajot}, {Paladini}, {Pasian}, {Patanchon},
  {Perdereau}, {Perotto}, {Perrotta}, {Piacentini}, {Piat}, {Pinheiro
  Gon{\c{c}}alves}, {Plaszczynski}, {Pointecouteau}, {Polenta}, {Ponthieu},
  {Poutanen}, {Pr{\'e}zeau}, {Prunet}, {Puget}, {Rachen}, {Reach}, {Reinecke},
  {Renault}, {Ricciardi}, {Riller}, {Ristorcelli}, {Rocha}, {Rosset},
  {Rowan-Robinson}, {Rubi{\~n}o-Mart{\'\i}n}, {Rusholme}, {Sandri}, {Santos},
  {Savini}, {Scott}, {Seiffert}, {Shellard}, {Smoot}, {Starck}, {Stivoli},
  {Stolyarov}, {Stompor}, {Sudiwala}, {Sygnet}, {Tauber}, {Terenzi},
  {Toffolatti}, {Tomasi}, {Torre}, {Tristram}, {Tuovinen}, {Umana},
  {Valenziano}, {Vielva}, {Villa}, {Vittorio}, {Wade}, {Wandelt}, {Wilkinson},
  {Yvon}, {Zacchei}, \& {Zonca}}]{planck:11}
{Planck Collaboration}, {Abergel}, A., {Ade}, P.~A.~R., {et~al.} 2011, \aap,
  536, A24

\bibitem[{{Price}(2007)}]{price:07}
{Price}, D.~J. 2007, \pasa, 24, 159

\bibitem[{Price \& Federrath(2010)}]{price:10}
Price, D.~J. \& Federrath, C. 2010, Monthly Notices of the Royal Astronomical
  Society, 406, 1659

\bibitem[{{Price} {et~al.}(2011){Price}, {Federrath}, \& {Brunt}}]{price:11}
{Price}, D.~J., {Federrath}, C., \& {Brunt}, C.~M. 2011, \apjl, 727, L21

\bibitem[{Price \& Laibe(2015)}]{price:15}
Price, D.~J. \& Laibe, G. 2015, Monthly Notices of the Royal Astronomical
  Society, 451, 813

\bibitem[{{Price} {et~al.}(2018){Price}, {Wurster}, {Tricco}, {Nixon},
  {Toupin}, {Pettitt}, {Chan}, {Mentiplay}, {Laibe}, {Glover}, {Dobbs},
  {Nealon}, {Liptai}, {Worpel}, {Bonnerot}, {Dipierro}, {Ballabio}, {Ragusa},
  {Federrath}, {Iaconi}, {Reichardt}, {Forgan}, {Hutchison}, {Constantino},
  {Ayliffe}, {Hirsh}, \& {Lodato}}]{price:18}
{Price}, D.~J., {Wurster}, J., {Tricco}, T.~S., {et~al.} 2018, \pasa, 35, e031

\bibitem[{{Roman-Duval} {et~al.}(2011){Roman-Duval}, {Federrath}, {Brunt},
  {Heyer}, {Jackson}, \& {Klessen}}]{roman:11}
{Roman-Duval}, J., {Federrath}, C., {Brunt}, C., {et~al.} 2011, \apj, 740, 120

\bibitem[{{Sadavoy} {et~al.}(2018){Sadavoy}, {Myers}, {Stephens}, {Tobin},
  {Commer{\c{c}}on}, {Henning}, {Looney}, {Kwon}, {Segura-Cox}, \&
  {Harris}}]{sadavoy:18}
{Sadavoy}, S.~I., {Myers}, P.~C., {Stephens}, I.~W., {et~al.} 2018, \apj, 859,
  165

\bibitem[{{Schmidt} {et~al.}(2009){Schmidt}, {Federrath}, {Hupp}, {Kern}, \&
  {Niemeyer}}]{schmidt:09}
{Schmidt}, W., {Federrath}, C., {Hupp}, M., {Kern}, S., \& {Niemeyer}, J.~C.
  2009, \aap, 494, 127

\bibitem[{Schmidt {et~al.}(2006)Schmidt, Hillebrandt, \& Niemeyer}]{schmidt:06}
Schmidt, W., Hillebrandt, W., \& Niemeyer, J.~C. 2006, Computers and Fluids,
  35, 353

\bibitem[{{Steinacker} {et~al.}(2010){Steinacker}, {Pagani}, {Bacmann}, \&
  {Guieu}}]{steinacker:10}
{Steinacker}, J., {Pagani}, L., {Bacmann}, A., \& {Guieu}, S. 2010, \aap, 511,
  A9

\bibitem[{{Teyssier}(2002)}]{teyssier:02}
{Teyssier}, R. 2002, \aap, 385, 337

\bibitem[{{Teyssier} \& {Commer{\c{c}}on}(2019)}]{teyssier:19}
{Teyssier}, R. \& {Commer{\c{c}}on}, B. 2019, Frontiers in Astronomy and Space
  Sciences, 6, 51

\bibitem[{Toro(1999)}]{toro:99}
Toro, E.~F. 1999, {Riemann Solvers and Numerical Methods for Fluid Dynamics}
  (Berlin, Heidelberg: Springer Berlin Heidelberg)

\bibitem[{{Tricco} {et~al.}(2017){Tricco}, {Price}, \& {Laibe}}]{tricco:17}
{Tricco}, T.~S., {Price}, D.~J., \& {Laibe}, G. 2017, \mnras, 471, L52

\bibitem[{{Tsukamoto} {et~al.}(2021){Tsukamoto}, {Machida}, \&
  {Inutsuka}}]{tsukamoto:21}
{Tsukamoto}, Y., {Machida}, M.~N., \& {Inutsuka}, S.-i. 2021, \apjl, 920, L35

\bibitem[{Weingartner \& Draine(2001)}]{weingartner:01}
Weingartner, J. \& Draine, B. 2001, The Astrophysical Journal, 548, 296

\bibitem[{{Youdin} \& {Goodman}(2005)}]{youdin:05}
{Youdin}, A.~N. \& {Goodman}, J. 2005, \apj, 620, 459

\end{thebibliography}

\begin{appendix}

\section{Resolution convergence \label{Sec:app_res}}
In this Appendix, we test the resolution convergence of the monofluid models. We repeat our fiducial setup 256MRN with resolutions of  $128^3$ and $512^3$. These models include ten dust species with sizes ranging from 1 nm to $20~\mu$m, distributed according to the MRN power law. 

\begin{figure*}[thb]
        \includegraphics[width=1\textwidth]{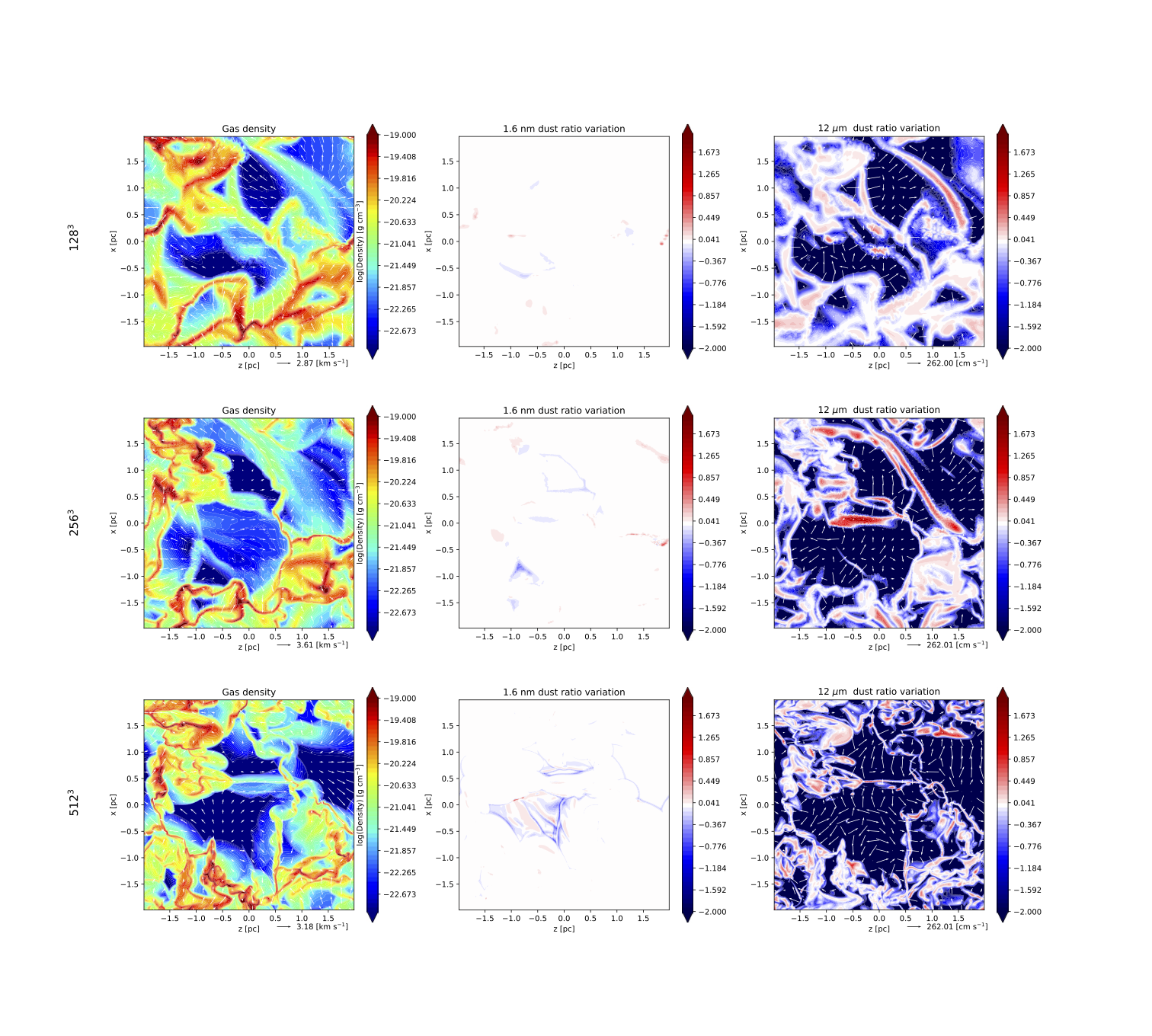}
        \caption{Total density (left), 1.6~nm dust-grain-ratio variations (middle), and 12~$\mu$m dust-grain-ratio variation  maps in the $xz$-plane for different resolutions: $128^3$ (top), $256^3$ (middle) and $512^3$ (bottom). The dust variation is given relative to the initial dust ratio value and is shown in logarithmic scale. The red colour shows dust-ratio enhancement while blue means a dust-ratio decrease. The arrows represent the barycentric velocity (left), and the 1.6~nm (middle) and 12~$\mu$m dust-grain (right) velocity vectors in the plane. All plots are made at a time corresponding to $2t_\mathrm{cross}.$
        }
        \label{Fig:resolution}
\end{figure*}

\begin{figure*}[thb]
        \includegraphics[width=1\textwidth]{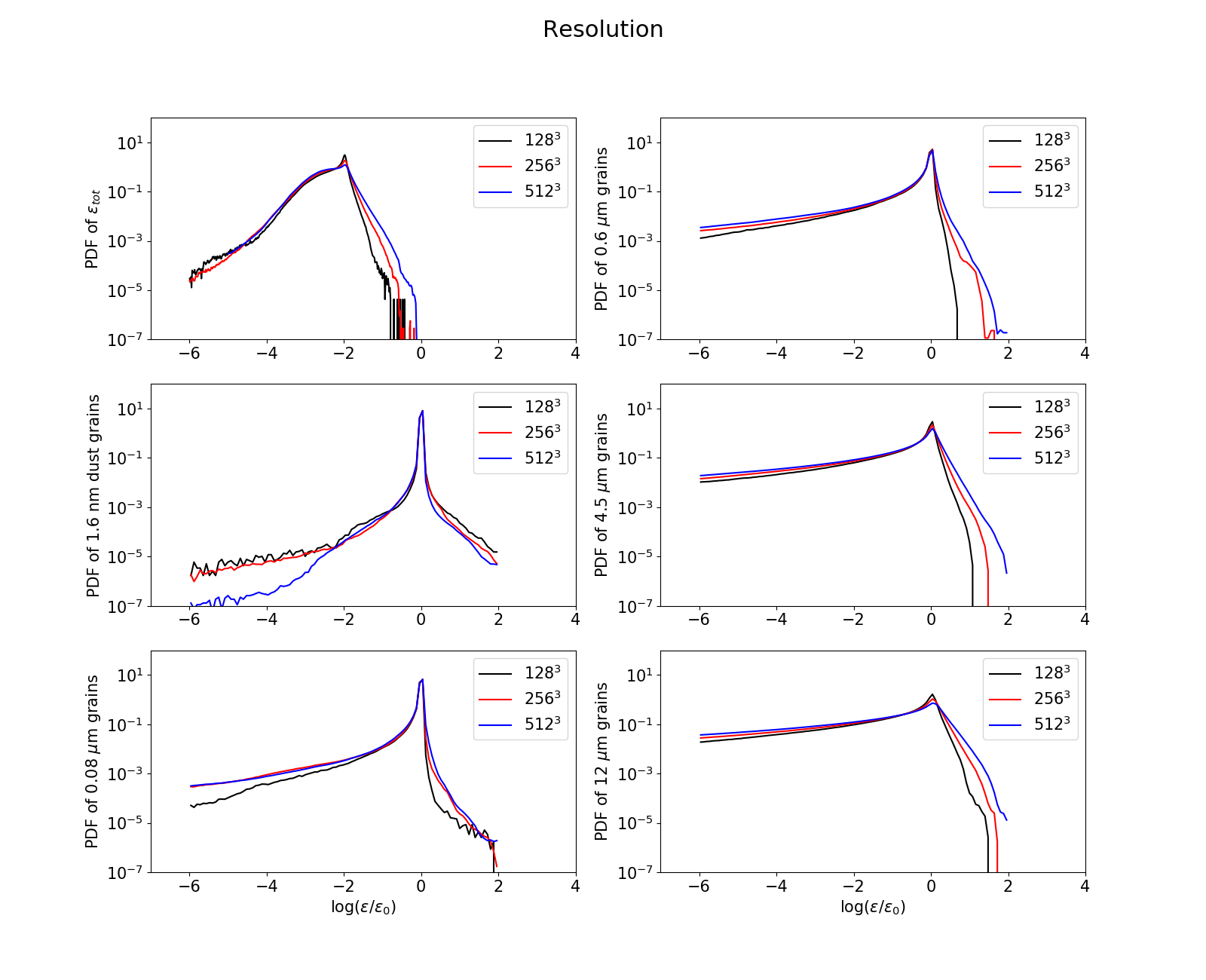}
        \caption{Dust-ratio PDFs for the total  dust-to-gas ratio, the 0.6~nm dust grains, the 1.6~nm dust grains, the 4.5~nm dust grains,  the 0.08~$\mu$m dust grains, and  the 12~$\mu$m dust grains. The black, red, and blue lines show respectively the $128$MRN, the $256$MRN, and the $512$MRN runs. The dust-ratio PDFs are normalised to the initial dust ratio of each dust species ($\epsilon_i/\epsilon_{i,0}$) and the PDFs of the total dust-to-gas ratio is not normalised to the initial dust ratio and thus peaks around 0.01.
        }
        \label{Fig:eps_resolution}
\end{figure*}

Figure \ref{Fig:resolution} shows the gas-density and dust-ratio variations of the two extreme dust-size bins. As expected in isothermal turbulence, the higher the resolution, the stronger the density contrasts. The dust-ratio variations are also sensitive to the resolution, with larger variations in the 512MRN model. 
Figure \ref{Fig:eps_resolution} portrays the  PDFs of the total dust-to-gas ratio and  of the dust ratios of the 1.6 nm, $0.08~\mu$m, $0.6~\mu$m, $4.5~\mu$m, and $12~\mu$m dust grains for the three resolutions. The PDFs get wider as resolution increases, except in the case of the 1.6~nm grains. Indeed, for the largest grains, the pressure gradients increase with resolution and thus the velocity shift between the dust and the gas. As a consequence, the dust grains tends to decouple more. For the smallest grains, the numerical diffusion decreases with resolution. The 1.6 nm grains should trace almost perfectly the gas since $\rm St <1$, see Fig. \ref{Fig:enrichement_histo}). As a consequence, as resolution increases, the coupling gets stronger.  

Our results are qualitatively very similar, and are independent of the resolution. The monofluid results we report in the main part of this study are thus robust  against resolution effects.

\section{Back-reaction\label{Sec:app_BR}}

Our two-fluid implementation does not include the back-reaction of the dust onto the gas. In order to compare the monofluid and two-fluid results, we neglected the back-reaction in the monofluid models presented in Sect.~\ref{Sec:bifluid}. To do so, we simply set the initial dust-to-gas ratio to $10^{-6}$ in the monofluid runs. In this Appendix, we compare the impact of the back-reaction in the monofluid models  in more depth than in the main
part of the manuscript. Our comparison is based on the 256MRN and on a simlar one but with an initial dust-to-gas ratio to $10^{-6}$

\begin{figure}[t]
        \includegraphics[width=0.5\textwidth]{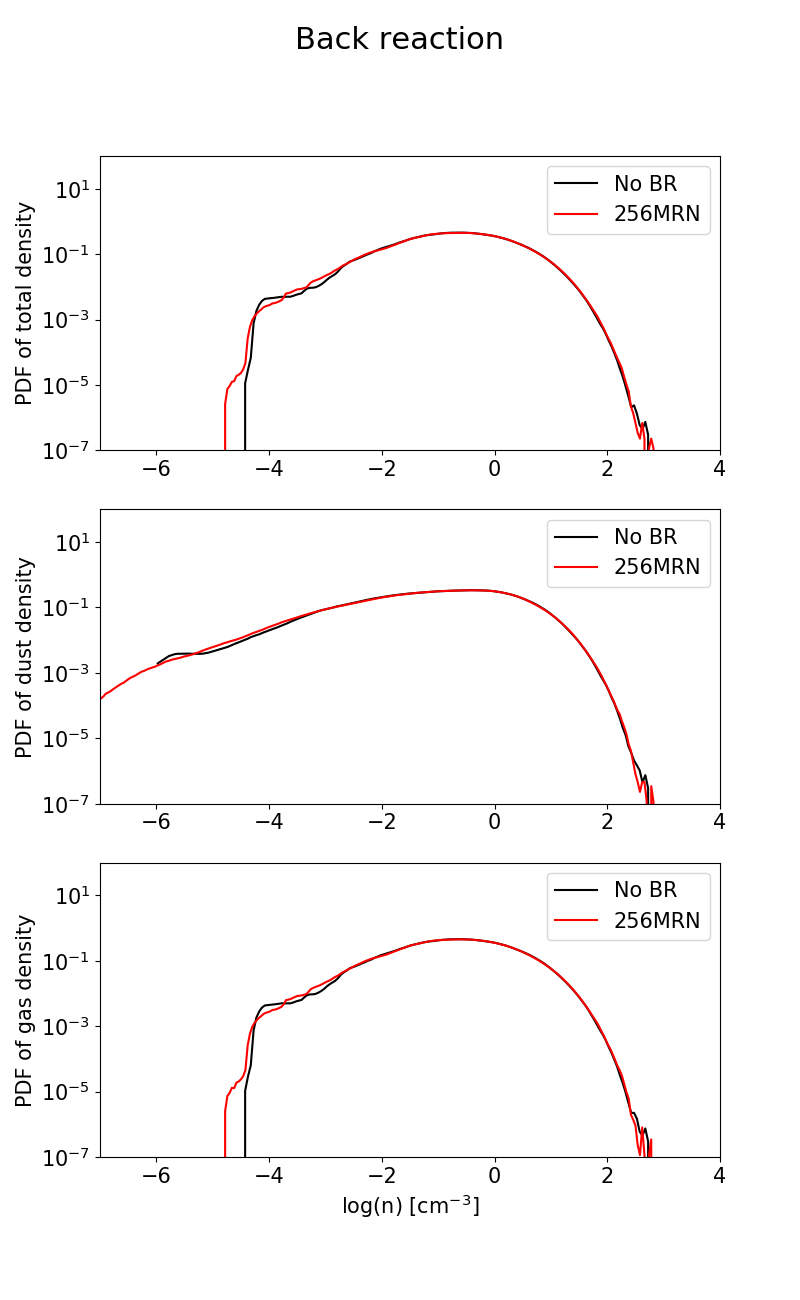}
        \caption{PDFs of the total (gas + dust) density (top), total dust density (middle), and gas density (bottom) variations for the 256MRN (fiducial, red) run and for the same model but with  an initial dust-to-gas ratio to $10^{-6}$ (black). All quantities are averaged over more than one crossing time. The different densities are normalised to their initial values. 
        }
        \label{Fig:pdf_backreaction}
\end{figure}

Figure~\ref{Fig:pdf_backreaction} shows the PDF of the total (gas + dust), gas, and total dust densities. First, the PDFs of the total and gas densities are almost identical, because the dust mass is always much less that the gas mass for the bulk of the material. The dust-enriched regions we report in the main text do not enclose sufficient mass to affect the global quantities. Second, all PDFs are very similar between the two runs. 

\begin{figure*}[t]
        \includegraphics[width=1\textwidth]{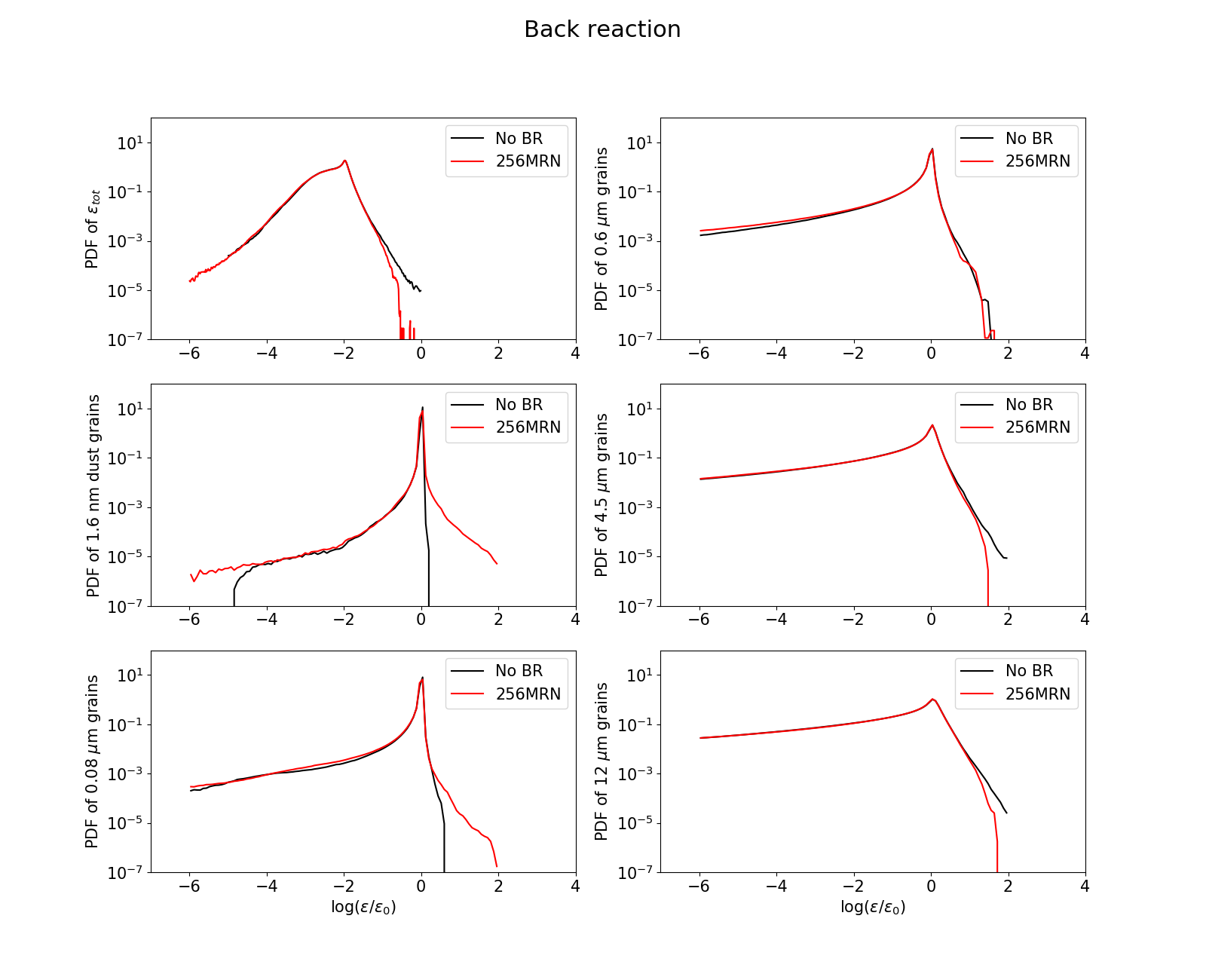}
        \caption{Same as Fig.~\ref{Fig:eps_resolution} for the comparison between the 256MRN run and the same one but with an initial dust-to-gas ratio of $10^{-6}$. 
        }
        \label{Fig:eps_backreaction}
\end{figure*}

Figure~\ref{Fig:eps_backreaction} shows the PDFs of the total dust-to-gas ratio and of the same dust ratios as in Fig.~\ref{Fig:eps_resolution} for the two runs. Only the tails of the PDFs are effected by the back-reaction, while the bulk of the mass remains unaffected. 
 
\end{appendix}

\end{document}